\documentclass[aps,prb,reprint,superscriptaddress]{revtex4-2}

\usepackage{amsmath, amssymb}
\usepackage{bm}
\usepackage{graphicx}
\usepackage{color}
\usepackage{dcolumn}
\usepackage{bm}
\usepackage[hidelinks]{hyperref}
\hypersetup{
  colorlinks   = true, 
  urlcolor     = blue, 
  linkcolor    = blue, 
  citecolor   = red 
}
\usepackage{multirow}

\allowdisplaybreaks[1]

\newcommand{\hH}{\hat{H}}

\begin{document}


\title{Vibronic excitations in resonant inelastic x-ray scattering spectra of K$_2$RuCl$_6$}

\author{Naoya Iwahara}
\email[]{naoya.iwahara@gmail.com}
\affiliation{Graduate School of Engineering, Chiba University, 1-33 Yayoi-cho, Inage-ku, Chiba-shi, Chiba 263-8522, Japan} 

\author{Shouta Shikano}
\affiliation{Department of Materials Science, Faculty of Engineering, Chiba University, 1-33 Yayoi-cho, Inage-ku, Chiba-shi, Chiba 263-8522, Japan}

\date{\today}

\begin{abstract}
We present the fingerprints of dynamic Jahn-Teller effect in resonant inelastic x-ray scattering (RIXS) spectra of K$_2$RuCl$_6$. 
We determined the dynamic Jahn-Teller model Hamiltonian of an embedded Ru$^{4+}$ ion using post Hartree-Fock methods, and derived the vibronic states by numerically diagonalizing the Hamiltonian. 
With the obtained vibronic states, we reproduced the RIXS spectra. 
The shape and the temperature dependence of the RIXS spectrum agree well with the experimental data. 
We found that some peaks emerge due to the dynamic Jahn-Teller effect rather than the crystal field splitting. 
Our study indicates the significance of the Jahn-Teller coupling to adequately interpret RIXS spectra. 
\end{abstract}

\maketitle

\section{Introduction}
Spin-orbit Mott insulators with heavy transition metal ions exhibit diverse quantum phenomena \cite{Witczak-Krempa2014, Rau2016, Takagi2019, Takayama2021}. 
A counterintuitive excitonic magnetic phase could emerge in compounds with nonmagnetic $t_{2g}^4$ ions \cite{Khaliullin2013}.
Heavy $t_{2g}^4$ ion embedded in an octahedral environment has a nonmagnetic $J=0$ ground state induced by strong spin-orbit coupling, whereas sufficiently strong exchange interaction between neighboring ions mixes the $J=0$ and excited magnetic $J=1$ multiplet states, and the admixed magnetic quantum states may condensate. 
The excitonic magnetism was attributed to the origin of the antiferromagnetism in Ca$_2$RuO$_4$ with a corner-shared structure.
In the excitonic magnetic phase close to the quantum critical point, amplitude fluctuation of magnetic moments (Higgs mode) develops, which was indeed observed in the spin-wave excitation of the compound \cite{Jain2017, Souliou2017}.
This theory predicts that different types of magnetism emerge in other lattices with edge-shared octahedra; zigzag one-dimensional magnetic order and bosonic Kitaev spin liquid phase in honeycomb lattice \cite{Khaliullin2013, Chaloupka2019}. 

\begin{figure}[tb]
\begin{tabular}{lll}
(a) & ~ & (b) \\
  \raisebox{0.165\linewidth}{
 \multirow{3}{*}{
\includegraphics[width=0.6\linewidth, bb = 0 0 719 539]{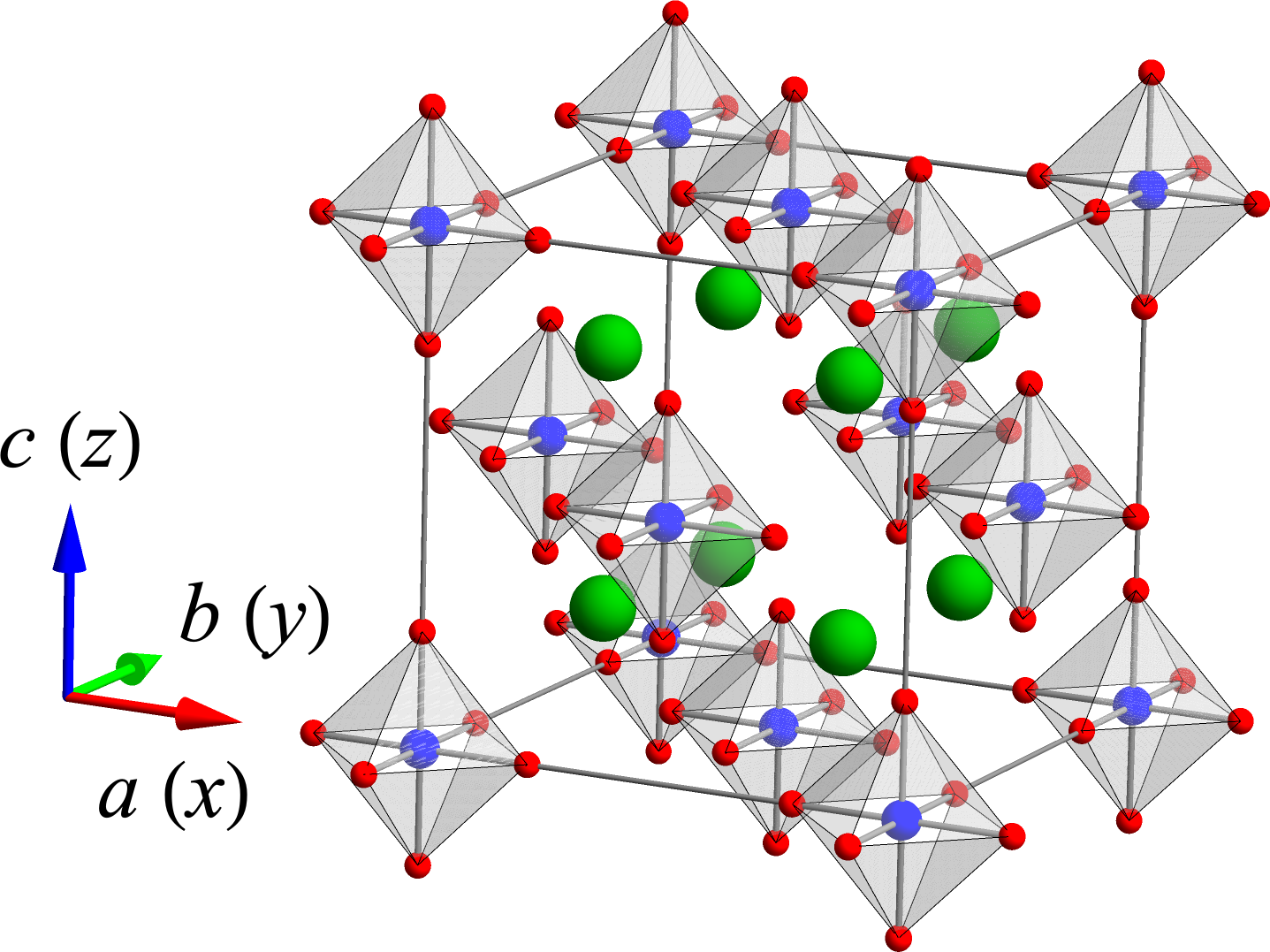}
  }
 }
 & & 
\includegraphics[width=0.20\linewidth, bb = 0 0 557 598]{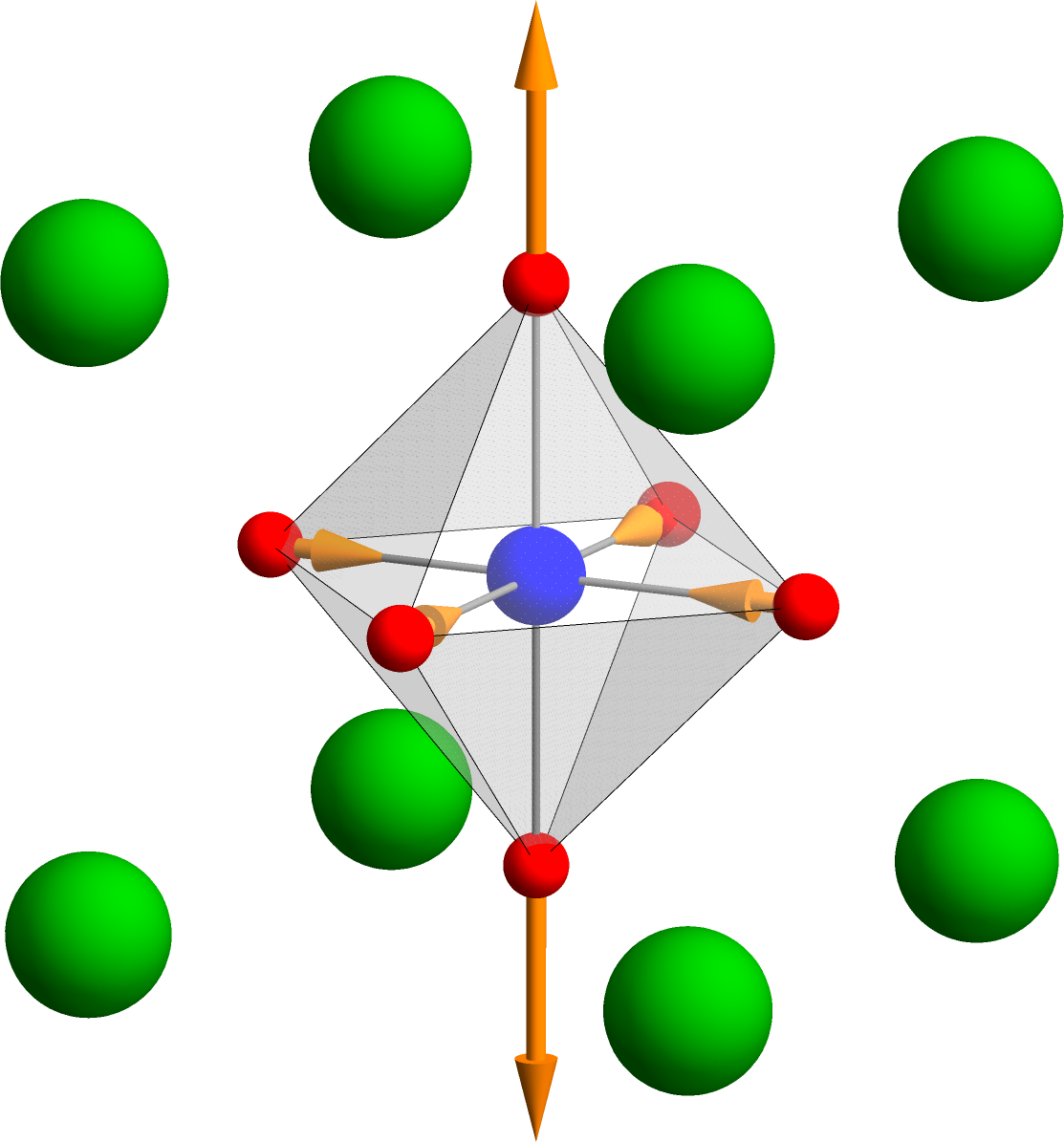}
\\
& & (c) \\
& & 
\includegraphics[width=0.20\linewidth, bb = 0 0 478 459]{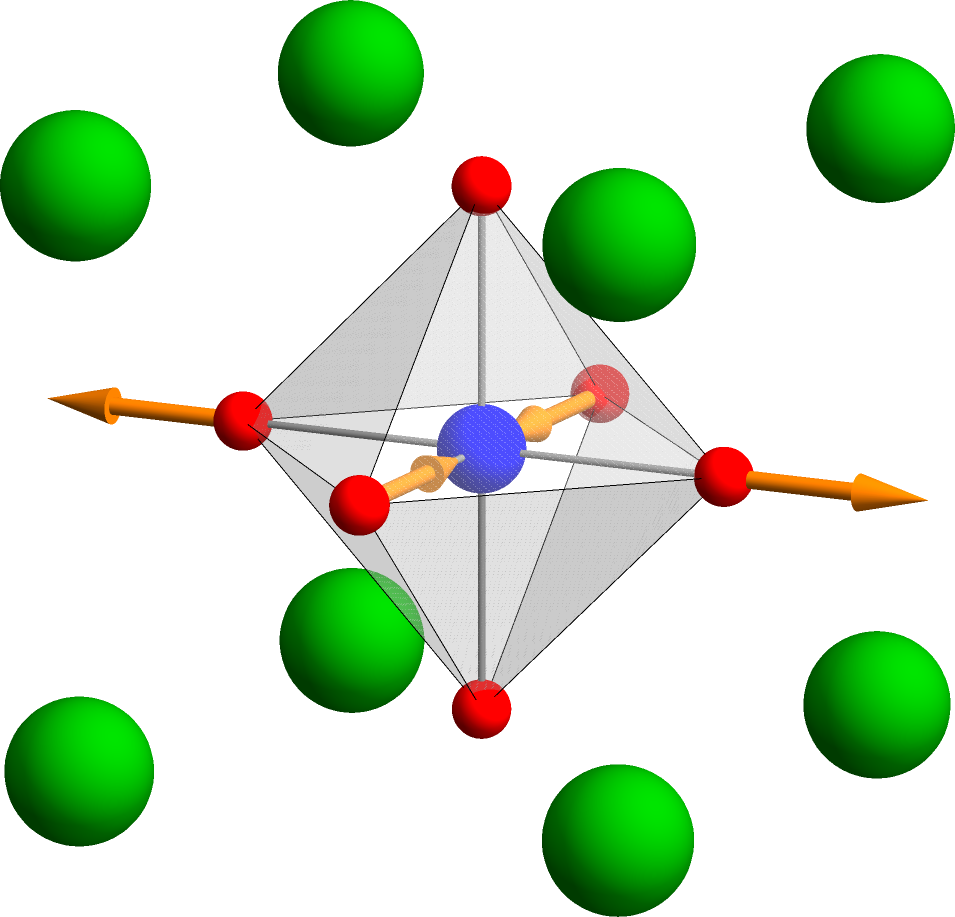}
\\
(d) \\
\multicolumn{3}{c}{
\includegraphics[width=0.90\linewidth, bb = 0 0 576 412]{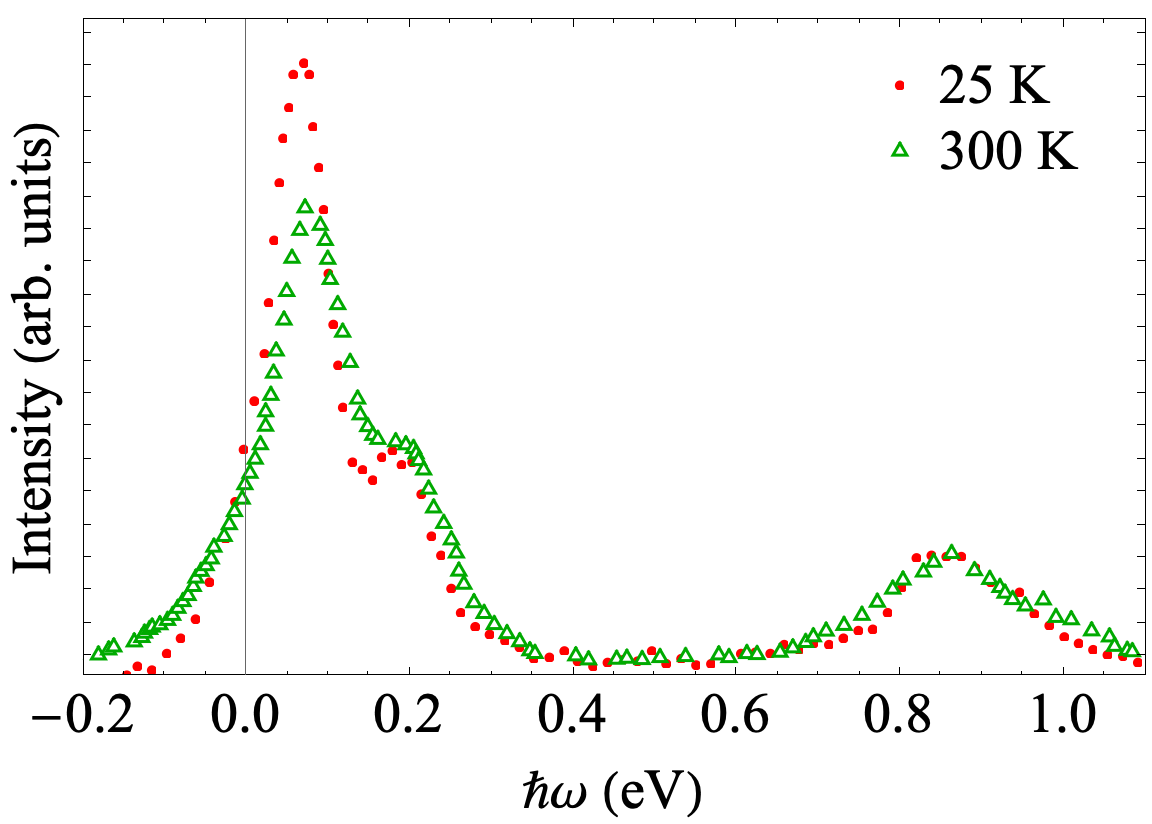}
}
\end{tabular}
\caption{
Crystal structure of K$_2$RuCl$_6$ and the experimental RIXS spectra. 
(a) Conventional cell of K$_2$RuCl$_6$. 
The blue, red, and green spheres are Ru, Cl, and K, respectively. 
The JT active (b) $E_gu$ ($3z^2-r^2$) and (c) $E_gv$ ($x^2-y^2$) modes. 
(d) Experimental RIXS spectra at 25 K (red circles) and 300 K (green open triangles) taken from Ref. \cite{Takahashi2021}.
We took the data with the momentum transfer of the light being $(hkl) = (0.19, 0.19, 0.19)$.
}
\label{Fig:k2rucl6}
\end{figure}

Experimental exploration of the excitonic magnetism in materials with edge-shared octahedra is underway.  
Early attempts towards the realization of the excitonic magnetism in Ir$^{5+}$ double perovskites were prevented by too strong spin-orbit coupling compared with the intersite exchange interaction \cite{Dey2016, Yuan2017, Fuchs2018}.
This situation lead researchers to investigate $4d^4$ compounds with weaker spin-orbit coupling than $5d^4$ compounds: the investigated compounds contain a honeycomb layered ruthenate, Ag$_3$LiRu$_2$O$_6$ \cite{Takayama2022}, and a cubic antifluorite, K$_2$RuCl$_6$ [Fig. \ref{Fig:k2rucl6}(a)] \cite{Takahashi2021}. 
In the former compound, three nonmagnetic phases arise under ambient and high-pressure conditions, while the excitonic magnetism does not develop \cite{Takayama2022}.
The latter is a Van Vleck-type diamagnetic material \cite{Hiraoka2021, Vishnoi2021}, whereas the exchange interaction between Ru sites is tiny in ambient pressure according to the dispersionless Ru $L_3$ resonant inelastic x-ray scattering (RIXS) spectra [Fig. \ref{Fig:k2rucl6}(d)] \cite{Takahashi2021}. 

An important factor controlling the magnetism in the $4d^4$ systems is the electron-phonon (vibronic) coupling. 
In Ca$_2$RuO$_4$, the vibronic coupling between the $J=0$ and the excited states causes the development of the pseudo JT deformation \cite{Bersuker1989, Bersuker2021}, which triggers the development of the spin-nematic phase above the magnetic transition \cite{Liu2019}.
In Ag$_3$LiRu$_2$O$_6$, the pseudo JT effect stabilizes a singlet dimer phase, preventing the emergence of excitonic magnetism \cite{Takayama2022}.

The vibronic coupling can give a significant influence on the energy spectrum of Ru ion in K$_2$RuCl$_6$ \cite{Takahashi2021}. 
In the compound, the spin-orbit coupling ($\lambda = 103 $ meV) extracted from the RIXS spectra is largely reduced compared with $\lambda = 167$ meV from the magnetic susceptibility data \cite{Vishnoi2021} and $\lambda = 150$ meV in $\alpha$-RuCl$_3$ \cite{Suzuki2021}.
Takahashi {\it et al}. attributed the large reduction of the spin-orbit coupling  to the dynamic JT stabilization of the $J=1$ states \cite{Takahashi2021}, while the magnitude of the vibronic coupling and the impact of the dynamic JT effect on RIXS spectra remain unclear. 

In this work, we prove the existence of the dynamic JT effect on the Ru sites in K$_2$RuCl$_6$ and elucidate its fingerprints in the RIXS spectra based on {\it ab initio} calculations. 
We derive a microscopic vibronic model of a Ru site with post Hartree-Fock calculations, and numerically diagonalize the vibronic Hamiltonian. 
With the obtained vibronic states, we simulate the RIXS spectra of K$_2$RuCl$_6$.

\section{Theory}
\subsection{Model Hamiltonian for $t_{2g}$ orbitals}
\label{Sec:t2g}
Let us set up our model for the RIXS in K$_2$RuCl$_6$. 
This compound is a face-centered cubic crystal consisting of RuCl$_6^{2-}$ octahedra [Fig. \ref{Fig:k2rucl6}(a)]. 
In each octahedron, ligand field splits the $4d$ orbitals into a doublet ($e_g$) and a triplet ($t_{2g}$), and the four $4d$ electrons populate the $t_{2g}$ orbitals \cite{Sugano1970}. 
On each site, the electrons feel Coulomb, spin-orbit, and electron-phonon (vibronic) couplings and the interplay of these interactions determines the local quantum states. 
The low-lying RIXS spectra display no variation with respect to the crystal momentum, suggesting that the intersite interactions between the neighboring octahedra are negligible \cite{Takahashi2021}. 

The model Hamiltonian for the embedded $t_{2g}^4$ ion consists of Coulomb $\hH_\text{C}$, spin-orbit $\hH_\text{SO}$ [\S 2.3.2 and \S 7.1.2 in Ref. \cite{Sugano1970}], vibronic $\hH_\text{JT}$ [\S 3.3 in Ref. \cite{Englman1972}, \S 3.3.2 in Ref. \cite{Bersuker1989}] interactions,
and harmonic oscillator Hamiltonian for the JT active modes $\hH_\text{vib}$:
\begin{align}
 \hH =& \hH_\text{C} + \hH_\text{SO} + \hH_\text{JT} + \hH_\text{vib},
 \label{Eq:H}
\\
 \hH_\text{C} =& 
 \sum_\gamma U \hat{n}_{\gamma\uparrow} \hat{n}_{\gamma\downarrow}
 +
 \sum_{\gamma < \gamma'} \sum_{\sigma\sigma'}
 (U - 2J_H) \hat{n}_{\gamma\sigma} \hat{n}_{\gamma'\sigma'}
 \nonumber\\
 &+
 \sum_{\gamma \ne \gamma'} J_H
 \hat{d}_{\gamma \uparrow}^\dagger 
 \hat{d}_{\gamma \downarrow}^\dagger 
 \hat{d}_{\gamma'\downarrow}
 \hat{d}_{\gamma'\uparrow}
 \nonumber\\
 &+
 \sum_{\gamma<\gamma'} \sum_{\sigma\sigma'} J_H
 \hat{d}_{\gamma\sigma}^\dagger
 \hat{d}_{\gamma'\sigma'}^\dagger
 \hat{d}_{\gamma\sigma'} 
 \hat{d}_{\gamma'\sigma},
 \label{Eq:HC}
 \\
 \hH_\text{SO} =& \sum_{\gamma\sigma} \sum_{\gamma'\sigma'} \lambda \langle \gamma\sigma| \hat{\bm{l}}\cdot \hat{\bm{s}} |\gamma'\sigma'\rangle \hat{d}_{\gamma\sigma}^\dagger \hat{d}_{\gamma'\sigma'},
 \label{Eq:HSO}
 \\
 \hH_\text{JT} =& \sum_{\sigma} V \left[
   \hat{n}_{yz,\sigma} \left(-\frac{1}{2} \hat{Q}_u + \frac{\sqrt{3}}{2} \hat{Q}_v \right)
   \right.
 \nonumber\\
 &+ 
 \left.
 \hat{n}_{zx,\sigma} \left(-\frac{1}{2} \hat{Q}_u - \frac{\sqrt{3}}{2} \hat{Q}_v \right)
 + \hat{n}_{xy,\sigma} \hat{Q}_u
 \right],
 \label{Eq:HJT}
 \\
 \hH_\text{vib} =&
 \sum_{\gamma=u,v} \frac{1}{2} \left(\hat{P}_\gamma^2 + \omega^2 \hat{Q}_\gamma^2\right).
 \label{Eq:Hvib}
\end{align}
Here $\hat{d}^\dagger_{\gamma\sigma}$ and $\hat{d}_{\gamma\sigma}$ are, respectively, electron creation and annihilation operators in orbital $\gamma$ ($\gamma = yz, zx, xy$) with spin projection $\sigma$, 
$\hat{n}_{\gamma\sigma} = \hat{d}^\dagger_{\gamma\sigma} \hat{d}_{\gamma\sigma}$ the electron number operator, 
$\hat{\bm{l}}$ the $t_{2g}$ orbital angular momenta [\S 7.1.1 in Ref. \cite{Sugano1970}], 
$\hat{\bm{s}}$ the spin angular momenta, 
$\hat{Q}_{\gamma}$ the mass-weighted normal coordinates [\S 10.1 in Ref. \cite{Inui1990}], 
$\hat{P}_\gamma$ the conjugate momenta, 
and $U$, $J_H$, $\lambda$, $V$, and $\omega$ are, respectively, the Coulomb, Hund's rule, spin-orbit, vibronic coupling parameters, and frequency. 
To obtain Eq. (\ref{Eq:HC}), we used the Slater-Condon integrals [\S 2.3.6 in Ref. \cite{Sugano1970}]. 
For the JT active modes, see Fig. \ref{Fig:k2rucl6}(b), (c).

Since the $t_{2g}$ orbitals are more than half-filled, we introduce hole operators. 
The hole creation $\tilde{d}^\dagger$ and annihilation $\tilde{d}$ operators are, respectively, 
\begin{align}
 \hat{d}^\dagger_{\gamma\sigma} = (-1)^{s+\sigma} \tilde{d}_{\gamma,-\sigma}, 
 \quad
 \hat{d}_{\gamma\sigma} = (-1)^{s+\sigma} \tilde{d}^\dagger_{\gamma,-\sigma}.
 \label{Eq:d_hole}
\end{align}
The vacuum state corresponds to the $t_{2g}^6$ electron configuration. 
The Coulomb interaction for the holes remains the same as Eq. (\ref{Eq:HC}) except for a constant term. 
We obtain it by replacing the $\hat{d}$ ($\hat{d}^\dagger$) with $\tilde{d}$ ($\tilde{d}^\dagger$) and using the constraint on the number of the holes per site, $\sum_{\gamma\sigma} \tilde{d}_{\gamma\sigma}^\dagger \tilde{d}_{\gamma\sigma} = \sum_{\gamma\sigma} \tilde{n}_{\gamma\sigma} = 2$. 
Similarly, by using Eq. (\ref{Eq:d_hole}), $\hat{H}_\text{SO}$ and $\hat{H}_\text{JT}$ remain the same form with opposite sign (Appendix \ref{A:hole}).

We also introduce dimensionless coordinates $\hat{q}$ and momenta $\hat{p}$:
\begin{align}
 \hat{Q}_{\gamma} = \sqrt{\frac{\hslash}{2\omega}} \hat{q}_\gamma, \quad 
 \hat{P}_{\gamma} = \sqrt{\frac{\hslash \omega}{2}} \hat{p}_\gamma.
\end{align}
With $\hat{q}$ and $\hat{p}$ and the hole operators (\ref{Eq:d_hole}), the vibronic coupling and harmonic oscillator Hamiltonian become, respectively,
\begin{align}
 \tilde{H}_\text{JT} =& \sum_{\sigma} -\hslash \omega g \left[
   \tilde{n}_{yz,\sigma} \left(-\frac{1}{2} \hat{q}_u + \frac{\sqrt{3}}{2} \hat{q}_v \right)
   \right.
 \nonumber\\
 &+ 
 \left.
 \tilde{n}_{zx,\sigma} \left(-\frac{1}{2} \hat{q}_u - \frac{\sqrt{3}}{2} \hat{q}_v \right)
 + \tilde{n}_{xy,\sigma} \hat{q}_u
 \right],
 \label{Eq:HJT_hole}
 \\
 \hH_\text{vib} =&
 \sum_{\gamma=u,v} \frac{\hslash \omega}{2} \left(\hat{p}_\gamma^2 + \hat{q}_\gamma^2\right).
 \label{Eq:Hvib_dimless}
\end{align}
Here $g$ stands for the dimensionless vibronic coupling parameter defined by 
\begin{align}
 g = \frac{V}{\sqrt{\hslash \omega^3}}.
 \label{Eq:g}
\end{align}

Now we diagonalize the interactions in the descending order of the energy scale. 
The Coulomb interaction splits the $t_{2g}^2$ hole configurations into four terms, $^3T_1 \oplus {}^1E \oplus {}^1T_2 \oplus {}^1A_1$ 
[The term energies are, respectively, $U-3J_H$, $U-J_H$, $U-J_H$, and $U+2J_H$. See \S 2.3.2 in Ref. \cite{Sugano1970}]. 
Since each representation appears once, we can uniquely determine the term states as 
\begin{align}
 |{}^{[S]}\Gamma \gamma M_S\rangle =& \frac{1}{\sqrt{2!}} 
 \sum_{\gamma_1\sigma_1} \sum_{\gamma_2\sigma_2} \tilde{d}_{\gamma_1\sigma_1}^\dagger \tilde{d}_{\gamma_2\sigma_2}^\dagger |0\rangle 
 \nonumber\\
 &\times
 (t_2\gamma_1, t_2\gamma_2|\Gamma\gamma) (s\sigma_1, s\sigma_2|SM_S),
 \label{Eq:SG}
\end{align}
where $(t_2\gamma_1, t_2\gamma_2|\Gamma\gamma)$ and $(s\sigma_1, s\sigma_2|SM_S)$ are the Clebsch-Gordan coefficients \cite{Koster1963, Varshalovich1988}, and $[S] = 2S+1$.
Using the term states (\ref{Eq:SG}) as the basis, the Coulomb Hamiltonian is
\begin{align}
 \hat{H}_\text{C} &= 
   2J_H  \left( \hat{P}_{{}^1E} + \hat{P}_{{}^1T_2} \right)
 + 5J_H \hat{P}_{{}^1A_1},
 \label{Eq:HH2}
\end{align}
with $\hat{P}_{{}^{[S]}\Gamma} = \sum_{\gamma M_S} |{}^{[S]}\Gamma \gamma M_S\rangle \langle {}^{[S]}\Gamma \gamma M_S|$.
In Eq. (\ref{Eq:HH2}), we set the ${}^3T_1$ term energy to zero. 

The spin-orbit coupling splits the $^{[S]}\Gamma$ terms into multiplet states. 
$\hat{H}_\text{SO}$ linearly couple to the $^3T_1$ term states, and the latter become $J = 0$ $(A_1)$, $J = 1$ $(T_1)$, and $J=2$ $(E \oplus T_2)$ multiplet states: 
\begin{align}
 |JM_M\rangle &= \sum_{\gamma M_S} |{}^3T_1\gamma M_S\rangle (t_1 \gamma, t_1 M_S|JM_J).
 \label{Eq:JM}
\end{align}
Using the $|JM_J\rangle$ and the spin singlet ${}^1\Gamma$ terms $(\Gamma = E, T_2, A_1$) as the basis for $\hat{H}_\text{SO}$, we obtain 
\begin{align}
\hat{H}_\text{SO} &= 
\lambda \Bigg[
 - \hat{P}_{J=0}
 - \frac{1}{2} \hat{P}_{J=1}
 + \frac{1}{2} \hat{P}_{J=2}
\nonumber\\
&
-i\sqrt{2}
 \left(|J=0\rangle \langle ^1A_1| - |^1A_1\rangle \langle J=0| \right)
\nonumber\\
&+
 \sum_{\Gamma=E,T_2} \sum_\gamma \frac{i}{\sqrt{2}} 
 \left( |\Gamma\gamma \rangle \langle {}^1\Gamma\gamma| - |{}^1\Gamma\gamma \rangle \langle \Gamma\gamma| \right)
\Bigg].
\label{Eq:HSO2}
\end{align}
Here $\hat{P}_J = \sum_{M_J=-J}^J |JM_J\rangle \langle JM_J|$.
The second and third lines in Eq. (\ref{Eq:HSO2}) are the interactions between the $J=0$ $(A_1)$ multiplet and ${}^1A_1$ term and between the $J=2$ $(E \oplus T_2)$ multiplet and ${}^1E \oplus {}^1T_2$ terms. 

The spin-orbit multiplet energy levels [the energy eigenstates of $\hat{H}_\text{C} + \hat{H}_\text{SO}$] are as follows: 
\begin{align}
 E_{A_1} &= \frac{1}{2} \Bigg(5 J_H - \lambda 
 - \sqrt{25J_H^2 + 10 J_H \lambda + 9 \lambda^2} \Bigg),
 \nonumber\\
 E_{T_1} &= - \frac{1}{2} \lambda, 
 \nonumber\\
 E_{E/T_2} &= \frac{1}{4} \left(4J_H + \lambda - \sqrt{16J_H^2 - 8J_H \lambda + 9\lambda^2}\right),
 \nonumber\\
 E_{{}^1E/{}^1T_2} &= \frac{1}{4} \left(4J_H + \lambda + \sqrt{16J_H^2 - 8J_H \lambda + 9\lambda^2}\right),
 \nonumber\\
 E_{{}^1A_1} &= \frac{1}{2} \Bigg(5 J_H - \lambda 
 + \sqrt{25J_H^2 + 10 J_H \lambda + 9 \lambda^2} \Bigg).
 \label{Eq:ESO}
\end{align}
The $^1E$, $^1T_2$, and $^1A_1$ states are no longer pure term states (\ref{Eq:SG}), while we continue using the same symbols. 

The JT interaction is active in orbitally degenerate terms. 
In the ${}^3T_1$ term, the orbital part of the vibronic interaction is 
\begin{align}
 &-\hslash \omega g
 \Bigg[
    \left(-\frac{1}{2} \hat{q}_u + \frac{\sqrt{3}}{2} \hat{q}_v \right) |{}^3T_1x\rangle \langle {}^3T_1x|
    \nonumber\\
    &+ \left(-\frac{1}{2} \hat{q}_u - \frac{\sqrt{3}}{2} \hat{q}_v \right) |{}^3T_1y\rangle \langle {}^3T_1y|
  + \hat{q}_u|{}^3T_1z\rangle \langle {}^3T_1z|
 \Bigg].
 \label{Eq:HJT3T1}
\end{align}
Transforming the ${}^3T_1$ term into the spin-orbit multiplet states (\ref{Eq:JM}), Eq. (\ref{Eq:HJT3T1}) reduces to 
\begin{widetext}
\begin{align}
-\hslash \omega g
\left(
\begin{array}{ccccccccc}
 0 & 0 & 0 & 0 & -\frac{1}{\sqrt{2}} \hat{q}_v & -\frac{1}{\sqrt{2}} \hat{q}_u & 0 & 0 & 0 \\
 0 & -\frac{1}{4} \hat{q}_u + \frac{\sqrt{3}}{4} \hat{q}_v & 0 & 0 & 0 & 0 & -\frac{3}{4} \hat{q}_u -\frac{\sqrt{3}}{4} \hat{q}_v & 0 & 0 \\
 0 & 0 & -\frac{1}{4} \hat{q}_u - \frac{\sqrt{3}}{4} \hat{q}_v & 0 & 0 & 0 & 0 & \frac{3}{4} \hat{q}_u -\frac{\sqrt{3}}{4} \hat{q}_v & 0 \\
 0 & 0 & 0 & \frac{1}{2} \hat{q}_u & 0 & 0 & 0 & 0 & \frac{\sqrt{3}}{2} \hat{q}_v  \\
 -\frac{1}{\sqrt{2}} \hat{q}_v & 0 & 0 & 0 & \frac{1}{2} \hat{q}_u & \frac{1}{2} \hat{q}_v & 0 & 0 & 0 \\
 -\frac{1}{\sqrt{2}} \hat{q}_u & 0 & 0 & 0 & \frac{1}{2} \hat{q}_v & -\frac{1}{2} \hat{q}_u & 0 & 0 & 0 \\
 0 & -\frac{3}{4} \hat{q}_u -\frac{\sqrt{3}}{4} \hat{q}_v & 0 & 0 & 0 & 0 & -\frac{1}{4} \hat{q}_u + \frac{\sqrt{3}}{4} \hat{q}_v & 0 & 0 \\
 0 & 0 & \frac{3}{4} \hat{q}_u -\frac{\sqrt{3}}{4} \hat{q}_v & 0 & 0 & 0 & 0 & -\frac{1}{4} \hat{q}_u - \frac{\sqrt{3}}{4} \hat{q}_v & 0 \\
 0 & 0 & 0 & \frac{\sqrt{3}}{2} \hat{q}_v & 0 & 0 & 0 & 0 & \frac{1}{2} \hat{q}_u 
\end{array}
\right),
 \label{Eq:HJTJ}
\end{align}
\end{widetext}
in the increasing order of $J$ [$J=0$, $J=1$ ($x,y,z$), $E$ ($u,v$) and $T_2$ ($yz, zx, xy$) from $J=2$]. 
Eq. (\ref{Eq:HJTJ}) consists of the $(A \oplus E) \otimes E$ and the $(T_1\oplus T_2) \otimes E$ JT interaction blocks. 
The diagonal blocks of Eq. (\ref{Eq:HJTJ}) indicate that the spin-orbit coupling quenches the vibronic coupling by half in comparison with Eq. (\ref{Eq:HJT_hole}).

The vibronic coupling is active within the ${}^1E \oplus {}^1T_2$ terms too. 
The JT Hamiltonian matrix for the terms is 
\begin{align}
-\hslash \omega g
 \left(
 \begin{array}{ccccc}
   \hat{q}_u & -\hat{q}_v & 0 & 0 & 0 \\
  -\hat{q}_v & -\hat{q}_u & 0 & 0 & 0 \\
     0 &    0 & \frac{1}{2} \hat{q}_u - \frac{\sqrt{3}}{2} \hat{q}_v & 0 & 0 \\
     0 &    0 & 0 & \frac{1}{2} \hat{q}_u + \frac{\sqrt{3}}{2} \hat{q}_v & 0 \\
     0 &    0 & 0 & 0 & -\hat{q}_u 
  \end{array}
 \right),
 \label{Eq:HJTET2}
\end{align}
in the order of ${}^1E$ $(u,v)$ and ${}^1T_2$ $(yz, zx, xy)$.
Eq. (\ref{Eq:HJTET2}) is the direct sum of $E \otimes E$ type and $T_2 \otimes E$ type JT interactions. 
The vibronic coupling in the $^1E \oplus {}^1T_2$ term states is unquenched. 

We ignore the vibronic coupling (\ref{Eq:HJT}) between different ${}^{[S]}\Gamma$ terms.
We show the validity of the assumption for K$_2$RuCl$_6$ in Sec. \ref{Sec:vibronic_abinitio}. 

The vibronic coupling of the JT type can drive the formation of the quantum entanglement of the spin-orbit multiplet and the vibrational states (dynamic JT effect). 
The energy eigenstates (vibronic states) of Eq. (\ref{Eq:H}) generally have the form of 
\begin{align}
 |\nu\rangle &= \sum_{\Gamma\gamma} |\Gamma\gamma\rangle \otimes |\chi_{\Gamma\gamma, \nu}\rangle,
 \label{Eq:vibronic}
\end{align}
where $|\Gamma\gamma\rangle$ indicate the spin-orbit multiplets, and $|\chi\rangle$ are the vibrational states of the JT modes. 
We determine $|\chi\rangle$ by a numerical method (Sec. \ref{Sec:vibronic}).
With the vibronic states (\ref{Eq:vibronic}) as the basis, the Hamiltonian is 
\begin{align}
\hat{H} = \sum_\nu E_\nu |\nu\rangle \langle \nu|,
\label{Eq:H0}
\end{align}
where $E_\nu$ are the energy eigenvalues.

\subsection{RIXS}
\label{Sec:RIXS_theory}
Here we describe the cross section for the Ru-$L_3$ RIXS taking account of the dynamic JT effect. 
The process consists of two steps: Excitation of an electron from the $2p_{3/2}$ orbitals to the empty $4d$ ($t_{2g}$) orbitals absorbing a photon followed by a transition of a $4d, t_{2g}$ electron into the empty $2p_{3/2}$ emitting a photon.
We can derive the cross section for the dynamic JT system by combining the vibronic states and the second order time-dependent perturbation theory [Kramers-Heisenberg formula. See \S 2.5 in Ref. \cite{Sakurai1967}].

The free Hamiltonian consists of the valence and core electron Hamiltonians and the radiation field Hamiltonian. 
We have described the valence Hamiltonian (\ref{Eq:H0}) in Sec. \ref{Sec:t2g}. 
The core level Hamiltonian is 
\begin{align}
 \hat{H}_c &= \sum_{m_j = -j}^{j} \epsilon_j \hat{c}_{jm_j}^\dagger \hat{c}_{jm_j},
\end{align}
where $j=\frac{3}{2}$, and $\hat{c}_{jm_j}^\dagger$ and $\hat{c}_{jm_j}$ are the electron creation and annihilation operators in $2p_{3/2}$ orbital with projection $m_j$: 
\begin{align}
 |jm_j \rangle &= \sum_{\gamma_p \sigma} |2p \gamma_p, s\sigma\rangle (\Gamma_4\gamma_p, \Gamma_6\sigma|\Gamma_8 m_j).
 \label{Eq:jm}
\end{align}
Here $\gamma_p = x, y, z$ are the components of the $2p$ orbitals. 
Single electron spin states and $j=\frac{3}{2}$ states belong to the $\Gamma_6$ and $\Gamma_8$ representations in the octahedron, respectively \cite{Koster1963}.  

The radiation field Hamiltonian is 
\begin{align}
 \hat{H}_\text{rad} &= \sum_{\bm{k}\lambda} \hslash \omega_k \left(\hat{a}_{\bm{k}\lambda}^\dagger \hat{a}_{\bm{k}\lambda} + \frac{1}{2} \right).
\end{align}
Here $\bm{k}$ are the momenta, $\lambda$ the polarization, $\omega_k = ck$ is the frequency of light, $k = |\bm{k}|$, and $c$ the speed of light.
$\hat{a}_{\bm{k}\lambda}^\dagger$ and $\hat{a}_{\bm{k}\lambda}$ are the creation and annihilation operators of the photon with $\bm{k}\lambda$, respectively.
With $\hat{a}_{\bm{k}\lambda}^\dagger$ and $\hat{a}_{\bm{k}\lambda}$, vector potential $\hat{\bm{A}}$ at the Ru site ($\bm{r}=\bm{0}$) is 
\begin{align}
 \hat{\bm{A}} &= \sum_{\bm{k}\lambda} \sqrt{\frac{\hslash}{2V\varepsilon_0\omega_{\bm{k}}}} 
 \left(
  \bm{e}_{\bm{k}\lambda} \hat{a}_{\bm{k}\lambda} 
  +
  \bm{e}_{-\bm{k}\lambda}^* \hat{a}_{-\bm{k}\lambda}^\dagger 
 \right).
 \label{Eq:A}
\end{align}
Here $\bm{e}_{\bm{k}\lambda}$ are the polarization vectors, $V$ is the volume, and $\varepsilon_0$ is the permittivity of the vacuum.
We take Coulomb gauge, and hence, $\bm{k}\cdot \bm{e}_{\bm{k}\lambda} = 0$.

We assume that the bilinear interaction of the $4d$ electron's momentum and the vector field be dominant and the field around the Ru site be uniform (dipole approximation):
\begin{align}
 \hat{H}' &= \frac{e}{m} \hat{\bm{A}} \cdot \hat{\bm{p}}.
 \label{Eq:Hdipole}
\end{align}
Here $e$ $(>0)$ is the elementary charge, $m$ is the mass of an electron, and $\hat{\bm{p}}$ is the momentum operator. 
The electron momentum operator between the core and valence orbitals is 
\begin{align}
\hat{\bm{p}}
&=
 \sum_{\gamma\sigma}
 \sum_{m_j}
 \left(
 \langle t_{2g}\gamma, s\sigma| \hat{\bm{p}} |jm_j\rangle \hat{d}_{\gamma\sigma}^\dagger \hat{c}_{jm_j} 
 \right.
 \nonumber\\
 &+
 \left.
 \langle jm_j| \hat{\bm{p}} |t_{2g}\gamma, s\sigma\rangle 
 \hat{c}_{jm_j}^\dagger 
 \hat{d}_{\gamma\sigma}
 \right).
 \label{Eq:p}
\end{align}
The matrix elements of $\hat{\bm{p}}$ are, by using Eq. (\ref{Eq:jm}) and Wigner-Eckart theorem \cite{Sugano1970, Inui1990, Koster1963, Varshalovich1988},
\begin{align}
 \langle t_{2g}\gamma, s\sigma| \hat{p}_\alpha |jm_j\rangle 
 &=
 \frac{(t_{2g}\Vert \hat{p} \Vert 2p)}{\sqrt{d_{t_{2}}}}
 \sum_{\gamma_p} (\Gamma_4\gamma_p, \Gamma_6 \sigma|\Gamma_8 m_j)
 \nonumber\\
 &\times
 (\Gamma_5 \gamma|\Gamma_4 \gamma_p, \Gamma_4\alpha). 
\end{align}
Here $\alpha = x, y, z$, 
$d_{t_2}=3$ is the dimension of the $t_2$ ($\Gamma_5$) representation, and $(t_{2g}\Vert \hat{p} \Vert 2p)$ is the reduced matrix element.

Applying the second-order time-dependent perturbation theory to our model under resonant condition, we obtain the cross section of the RIXS processes. 
The initial and final states are the products of the vibronic states (\ref{Eq:vibronic}) and one-photon states, $|\bm{k}\lambda\rangle = \hat{a}_{\bm{k}\lambda}^\dagger|0\rangle$, and the intermediate states are those with one $2p_{3/2}$ core-hole. 
When the initial and intermediate energies are close to each other, the cross section is 
\begin{align}
 \frac{d^2\sigma}{d\Omega dk'} =& 
 \frac{V^2\omega_{k'}^2}{(2\pi)^2\hslash c^3}
 \left|
  \langle \nu'; \bm{k}'\lambda'|
   \hat{H}' \hat{G}(z_{\nu k}) 
   \hat{H}'
  |\nu; \bm{k}\lambda \rangle 
 \right|^2
 \nonumber\\
 &\times
 \delta\left(E_\nu + \hslash \omega_k - E_{\nu'} - \hslash \omega_{k'}\right).
 \label{Eq:cross}
\end{align}
Here $\hat{G}$ is the propagator for the intermediate states, $\hat{G}(z) = \sum_n \frac{|n\rangle \langle n|}{z - E_n}$, and $z_{\nu k} = E_\nu + \hslash \omega_k + i \Gamma$.
Substituting $\hat{H'}$ (\ref{Eq:Hdipole}) into Eq. (\ref{Eq:cross}), we obtain an explicit form for the vibronic RIXS spectrum:
\begin{align}
 \frac{d^2\sigma}{d\Omega dk'} =& 
 \frac{\hslash c a_0^2}{m^2} \frac{\omega_{k'}}{\omega_k}
 \left|
 \sum_{\alpha\alpha'}
  e_{\bm{k}'\lambda', \alpha'}^*
  e_{\bm{k}\lambda, \alpha}
  \langle \nu'|
   \hat{p}_{\alpha'}
   \hat{G}(z_{\nu k})
   \hat{p}_{\alpha}
  |\nu\rangle
 \right|^2
\nonumber\\
&\times
 \delta
 \left(
  E_\nu + \hslash \omega_k - E_{\nu'} - \hslash \omega_{k'}
 \right).
 \label{Eq:cross_1}
\end{align}
The vibronic cross-section indicates that the dynamic JT effect modulates the RIXS spectrum in two ways:
(1) the vibronic reduction of the electronic operator $\hat{p}_{\alpha'} \hat{G}(z_{\nu k}) \hat{p}_{\alpha}$ and (2) the emergence of new peaks.

We continue simplifying the cross section for our numerical calculations by applying the fast collision approximation \cite{Luo1993, vanVeenendaal2006}.
This approximation ignores the detailed energy structures and dynamics of the intermediate states by replacing $E_n$ and $\hat{G}$ by a typical value $\bar{E}$ and $\bar{G}(z) = (z - \bar{E})^{-1}$, respectively.
With the approximation, Eq. (\ref{Eq:cross_1}) reduces to 
\begin{align}
 \frac{d^2\sigma}{d\Omega dk'} =& 
 \frac{\hslash c a_0^2}{m^2} \frac{\omega_{k'}}{\omega_k}
 \left|\bar{G}(z_{\nu k})\right|^2
\nonumber\\
&\times
 \left|
 \sum_{\alpha\alpha'}
  e_{\bm{k}'\lambda', \alpha'}^*
  e_{\bm{k}\lambda, \alpha}
  \langle \nu'|
  \hat{F}_{\alpha'\alpha}
  |\nu\rangle
 \right|^2
\nonumber\\
&\times
 \delta
 \left(
  E_\nu + \hslash \omega_k - E_{\nu'} - \hslash \omega_{k'}
 \right),
 \label{Eq:cross_section_0K}
\end{align}
where $\hat{F}_{\alpha'\alpha} = \hat{p}_\alpha \hat{P}_\text{ch}\hat{p}_{\alpha'}$, and $\hat{P}_\text{ch}$ the projection operator into the intermediate core-hole states. 
Using Eq. (\ref{Eq:p}) in $\hat{F}$, 
\begin{align}
 \hat{F}_{\alpha'\alpha} &= 
 \frac{(t_{2g} \Vert \hat{p} \Vert 2p)^2}{d_{t_2}}
 \sum_{\gamma'\sigma'}\sum_{\gamma\sigma}
  (-1)^{\sigma-\sigma'}
\nonumber\\
&\times
 \Bigg[
 \sum_{\gamma_p'\gamma_p} 
 \sum_{m_j}
 (t_1\gamma_p', \Gamma_6 \sigma'|\Gamma_8 m_j)^*
 (t_1\gamma_p', t_1\alpha'|t_2\gamma')
\nonumber\\
&\times
 (t_1\gamma_p, \Gamma_6 \sigma|\Gamma_8 m_j)
 (t_1\gamma_p, t_1\alpha|t_2\gamma)
 \Bigg]
  \tilde{d}_{\gamma'-\sigma'}^\dagger 
  \tilde{d}_{\gamma,-\sigma}.
\end{align}

Finally, we include the thermal effect.
The cross-section at finite temperature is 
\begin{align}
 \frac{d^2 \sigma}{d\Omega dk'} &= 
 \frac{\hslash c a_0^2}{m^2} \frac{\omega_{k'}}{\omega_k}
 \left|\bar{G}(z_{\nu k})\right|^2
 \nonumber\\
 &\times 
 \sum_\nu 
 \rho_\nu 
 \left|
 \sum_{\alpha\alpha'}
  e_{\bm{k}'\lambda', \alpha'}^*
  e_{\bm{k}\lambda, \alpha}
  \langle \nu'|
  \hat{F}_{\alpha'\alpha}
  |\nu\rangle
 \right|^2
\nonumber\\
&\times
 \delta
 \left(
  E_\nu + \hslash \omega_k - E_{\nu'} - \hslash \omega_{k'}
 \right),
 \label{Eq:cross_section_T}
\end{align}
with the canonical distribution of the dynamic JT system, $\rho_\nu = \exp(-E_\nu \beta)/Z$. 
Here $\beta$ is the inverse temperature and $Z = \sum_\nu \exp(-E_\nu \beta)$.

\section{Methods}
\label{Sec:method}
\subsection{{\it Ab initio} method}
\label{Sec:abinitio}
We quantitatively determined the electronic structure of a single Ru site by cluster calculations with post Hartree-Fock methods. 
We constructed the Ru cluster from the x-ray structure at 300 K \cite{Vishnoi2021}.
The cluster consists of three parts. 
The first part contains one Ru atom, the nearest six Cl, and the nearest eight K atoms.
We treated the electrons in this part fully quantum mechanically with the atomic-natural-orbital relativistic-correlation consistent-valence triple zeta polarization (ANO-RCC-VTZP) basis functions.
The second part consists of surrounding atoms (12 Zr atoms at the Ru sites, 48 K, and 72 Cl). 
We treated them within the {\it ab initio} embedding model potential method \cite{Seijo1999}.
The last part consists of 1554 point charges surrounding the first and the second parts. 
The total charge of the cluster is neutral.

We calculated the electronic states of the cluster employing a series of post Hartree-Fock methods. 
First, we derived the ${}^{[S]}\Gamma$ term states using the complete active space self-consistent field (CASSCF) method \cite{Roos2016}.
In the CASSCF calculations, we treated the five $4d$ orbitals as the active space and calculated all the term states with $S = 0, 1, 2$.
We expressed the atomic bielectronic integrals using Cholesky decomposition with a threshold of $5 \times 10^{-7}$ $E_h$ and set the ionization potential electron affinity (IPEA) shift to zero and the imaginary (IMAG) shift to 0.1.
After the CASSCF calculations, we included the dynamic electron correction effect on the ${}^{[S]}\Gamma$ term energies with the extended multistate complete active space second-order perturbation theory (XMS-CASPT2) \cite{Granovsky2011, Shiozaki2011}.
Then, we included the spin-orbit coupling using the spin-orbit restricted active space state interaction (SO-RASSI) method. 
For all the calculations, we used {\tt OpenMolcas} \cite{molcas1, molcas2}.

\subsection{Vibronic coupling parameters}
\label{Sec:vibronic_coupling}
We derived the vibronic coupling parameters by fitting the ${}^3T_1$ energy levels for JT deformed structures to the JT model as in Refs. \cite{Iwahara2017, Iwahara2018}. 
We constructed the JT deformed structures of the Ru cluster by varying the normal coordinate $Q_u$ from 0 to 10 by 2 (in atomic unit):
\begin{align}
 \bm{R}_A = \bm{R}_A^{(0)} + \frac{Q_u}{\sqrt{M_A}} \left( \bm{e}_{u} \right)_A.
 \label{Eq:RA}
\end{align}
Here $A$ indicates the nearest neighbor Cl atoms, $\bm{R}_A$ the Cartesian coordinates of atom $A$, $\bm{R}_A^{(0)}$ the coordinates at the perfect octahedral structure, $M_A$ the mass of atom $A$, $\bm{e}_u$ the eigenvector of the dynamical matrix, and $\left( \bm{e}_u \right)_A$ the components for atom $A$ in $\bm{e}_u$. 
We chose the phase of $\bm{e}_u$ to give the deformation in Fig. \ref{Fig:k2rucl6}(b) with positive $Q_u$.
At each JT deformed structure, we performed the CASSCF/XMS-CASPT2 calculations. 

We obtained $\omega$ and the vibronic coupling parameter $g$ by fitting the {\it ab initio} term energies to the potential energy surface of the JT model. 
The model potential contains the harmonic potential and the vibronic coupling (\ref{Eq:HJT3T1}):
\begin{align}
 U(Q_u) &= \frac{\omega^2}{2} Q_u^2 
    -V
    \left(-\frac{1}{2} Q_u |{}^3T_1x\rangle \langle {}^3T_1x|
    \right.
\nonumber\\
    &-
    \left.
    \frac{1}{2} Q_u |{}^3T_1y\rangle \langle {}^3T_1y|
  + Q_u|{}^3T_1z\rangle \langle {}^3T_1z|
 \right).
 \label{Eq:U}
\end{align}

\subsection{Vibronic states}
\label{Sec:vibronic}
We calculated the vibronic states by numerically diagonalizing the dynamic JT Hamiltonian (\ref{Eq:H}).
We expand the nuclear part $|\chi\rangle$ of the vibronic states (\ref{Eq:vibronic}) with the energy eigenstates of $\hat{H}_\text{vib}$, $|n_u, n_v\rangle$ ($n_u, n_v = 0, 1, 2, ...$), and expansion coefficients, $\chi_{\Gamma\gamma n_u n_v, \nu}$:
\begin{align}
 |\chi_{\Gamma\gamma, \nu} \rangle &= \sum_{n_u, n_v} |n_u, n_v\rangle \chi_{\Gamma\gamma n_u n_v, \nu}. 
 \label{Eq:chi}
\end{align}
Thus, the vibronic basis for the dynamic JT Hamiltonian is a set of the direct products of $|\Gamma\gamma\rangle \otimes |n_u, n_v\rangle$.

To numerically diagonalize the vibronic Hamiltonian, we introduced the following approximations. 
We treated the vibronic states related to the ${}^3T_1$ terms and the ${}^1E \oplus {}^1T_2$ terms separately. 
This is valid when the pseudo JT couplings between the terms are negligible. 
We truncated the vibronic basis by introducing the maximum number of the vibrational quanta, $0 \le n_u + n_v \le 20$. 
This basis is sufficiently large [See Ref. \cite{Iwahara2018}].

With the vibronic basis, we constructed the vibronic Hamiltonian matrix, and numerically diagonalized it. 
For the diagonalization of the Hamiltonian matrix, we used {\tt dsyevd} in {\tt Lapack} library \cite{laug}.

\section{Results} 
\label{Sec:results}

\begin{figure}[tb]
\begin{tabular}{lll}
 (a) &~& (b) \\
 \includegraphics[bb = 0 0 576 557, height=0.45\linewidth]{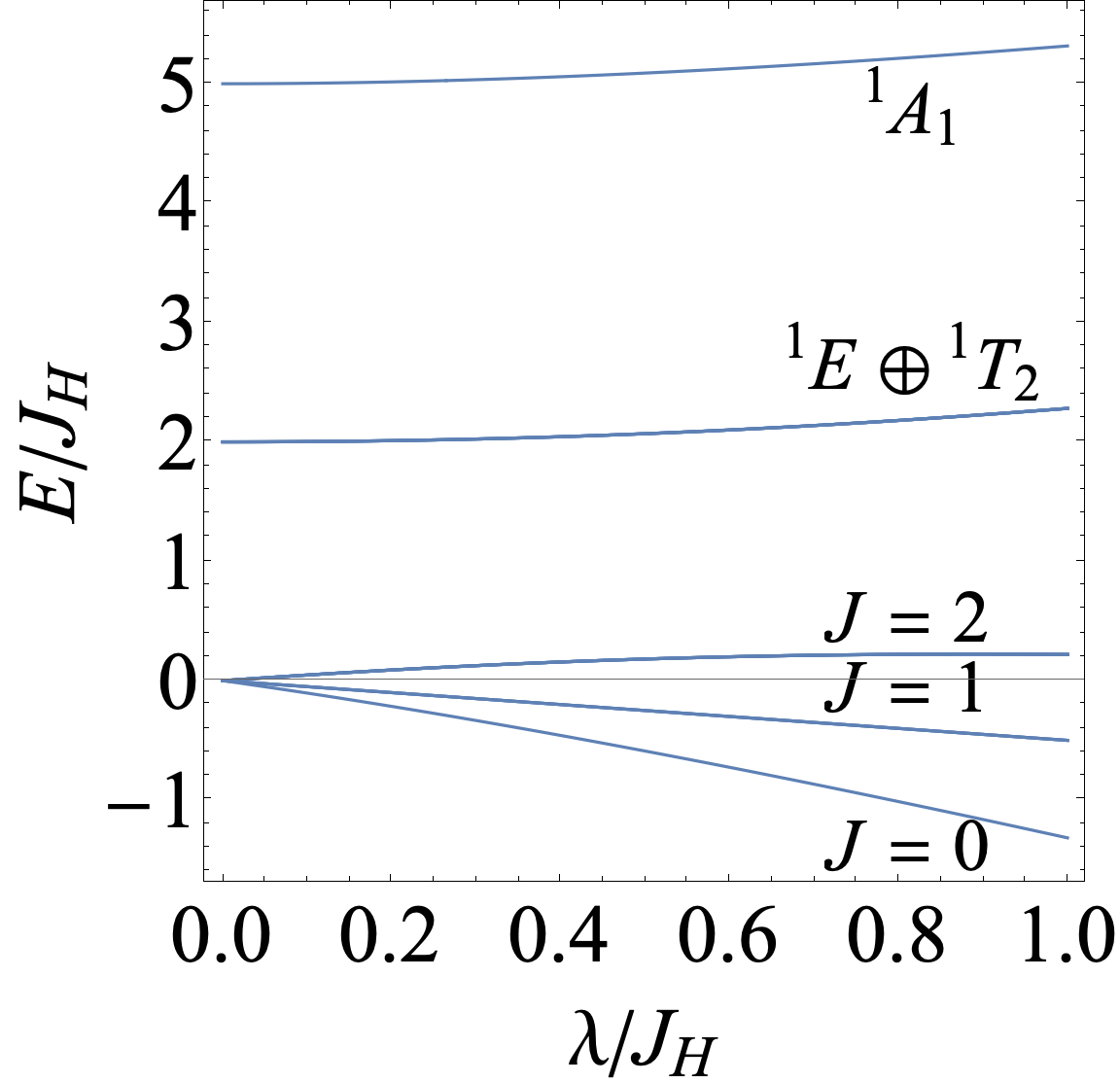}
 & &
 \includegraphics[bb = 0 0 576 540, height=0.45\linewidth]{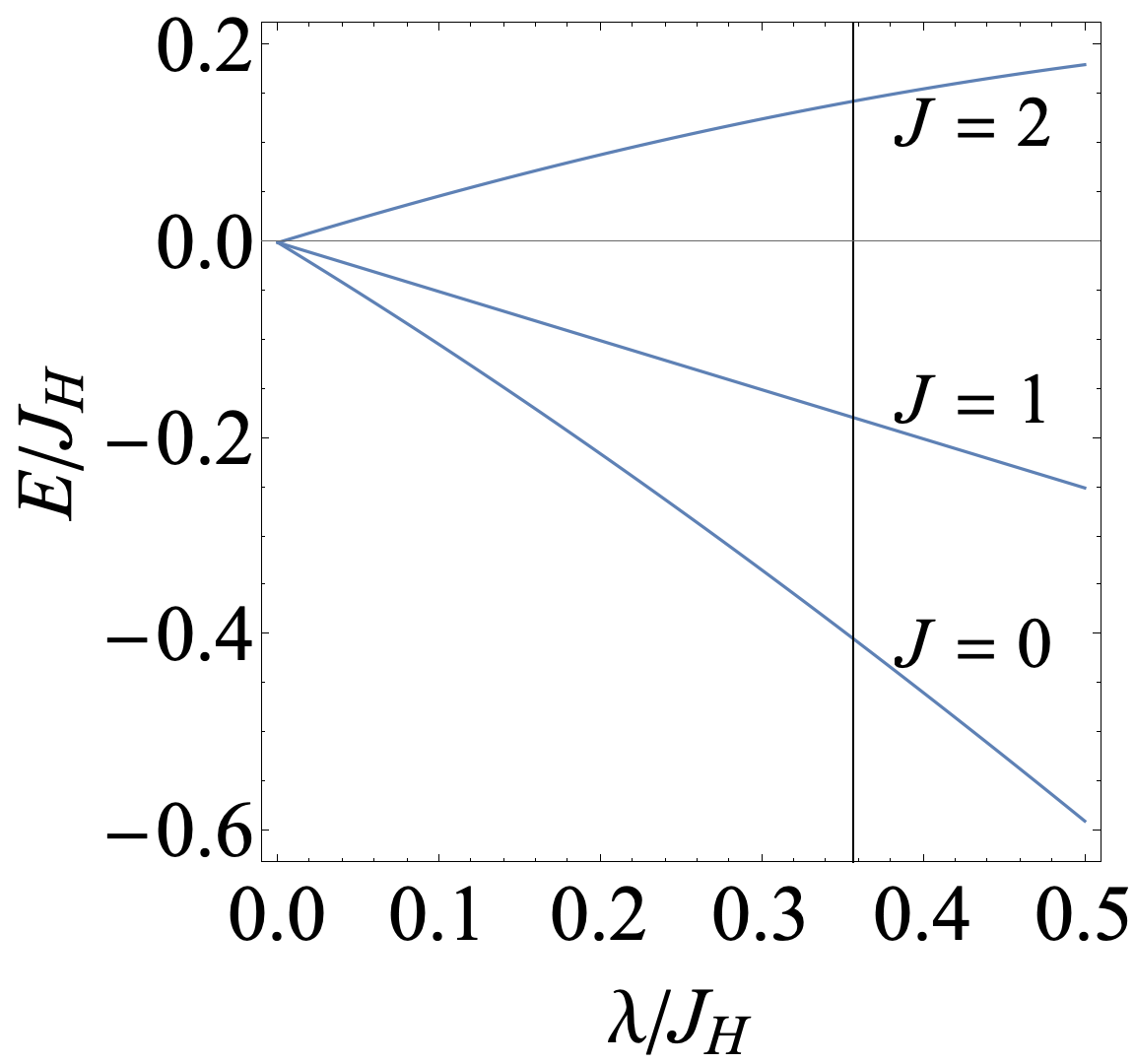}
\end{tabular}
\caption{
  The spin-orbit energy levels with respect to $\lambda/J_H$ for the $O_h$ cluster.
  The spin-orbit multiplet energies originate from (a) the $t_2^4$ states and (b) the $^3T_1$ term. 
  At the vertical line ($\lambda/J_H = 0.357$), the ratio of the excited energy levels with respect to the ground one coincides with the {\it ab initio} data. 
}
\label{Fig:SO}
\end{figure}

\subsection{Electronic states}
\label{Sec:electronic_abinitio}
We performed the {\it ab initio} electronic state calculations of the cubic Ru cluster. 
Table \ref{Table:E} shows the calculated electronic energy levels: the values of the left and right columns correspond to the ${}^{[S]}\Gamma$ term and spin-orbit multiplet energies, respectively. 
The splittings of the $J=2$ and the excited $E \otimes T_2$ multiplet energy levels amount to only a few meV. 

\begin{table}[tb]
\begin{ruledtabular}
\caption{{\it Ab initio} ${}^{[S]}\Gamma$ term and spin-orbit multiplet energies of the cubic Ru cluster (eV). 
}
\label{Table:E}
\begin{tabular}{llll}
\multicolumn{2}{c}{${}^{[S]}\Gamma$ term} & \multicolumn{2}{c}{Spin-orbit multiplet} \\ 
\hline
$^3T_1$ & 0      & $J=0~(A_1)$ & $-0.1604$ \\
        &        & $J=1~(T_1)$ & $-0.0756$ \\
        &        & $J=2~(T_2)$ &   0.0455  \\
        &        & $J=2~(E)$   &   0.0461  \\
$^1T_2$ & 0.9585 &             &   0.9581  \\
$^1E$   & 0.9629 &             &   0.9619  \\
$^1A_1$ & 2.1275 &             &   2.1350  \\
\end{tabular}
\end{ruledtabular}
\end{table}

We determined the electronic interaction parameters by fitting the {\it ab initio} data to the model Hamiltonians. 
We obtained $J_H =$ 443.7 meV from the fitting of the gaps of the CASSCF/XMS-CASPT2 levels to Eq. (\ref{Eq:HH2}). 
Since the energy splitting of the $^1T_2$ and $^1E$ is only 4 meV and much smaller than the other energy gaps, we ignored the splitting in the fitting. 
The present Hund rule coupling is close to the experimental $J_H = $ 420 meV extracted from the RIXS spectra of K$_2$RuCl$_6$ \cite{Takahashi2021}.

Similarly, we derived the spin-orbit coupling parameter from the SO-RASSI levels and Eq. (\ref{Eq:ESO}). 
The energy levels of the first ($J=1$) and the second ($J=2$) excited states with respect to the ground ($J=0$) level are, respectively, $\Delta E_{J=1} = $ 84.8 meV and $\Delta E_{J=2} =$ 206.1 meV ignoring the small splitting of the latter.
We determined $\lambda/J_H$ to be 0.357 by reproducing the ratio of $\Delta E_{J=2}/\Delta E_{J=1} = 2.4$ with Eq. (\ref{Eq:ESO}) [Fig. \ref{Fig:SO}]. 
Our spin-orbit coupling $\lambda$ is 158 meV. 

The {\it ab initio} $\lambda$ deviates from the experimental estimate in Ref. \cite{Takahashi2021}. 
The ratio of the excitation energies, $\Delta E_{J=2}/\Delta E_{J=1} = 2.4$, is smaller than the ratio of 2.7 extracted from the RIXS data.
The present $\lambda$ is close to $\lambda = 167$ meV derived from the magnetic susceptibility data of K$_2$RuCl$_6$ \cite{Vishnoi2021} and $\lambda = 150$ meV for $\alpha$-RuCl$_3$ \cite{Suzuki2021}, while by about 50 \% larger than $\lambda = 103$ meV derived from the RIXS spectra \cite{Takahashi2021}.
Takahashi {\it et al}. ascribed this reduction to the dynamic JT effect. 
We examine this idea below. 

\begin{figure}[tb]
\begin{tabular}{lll}
 (a) &~& (c) \\
 \includegraphics[bb = 0 0 518 489, height=0.46\linewidth]{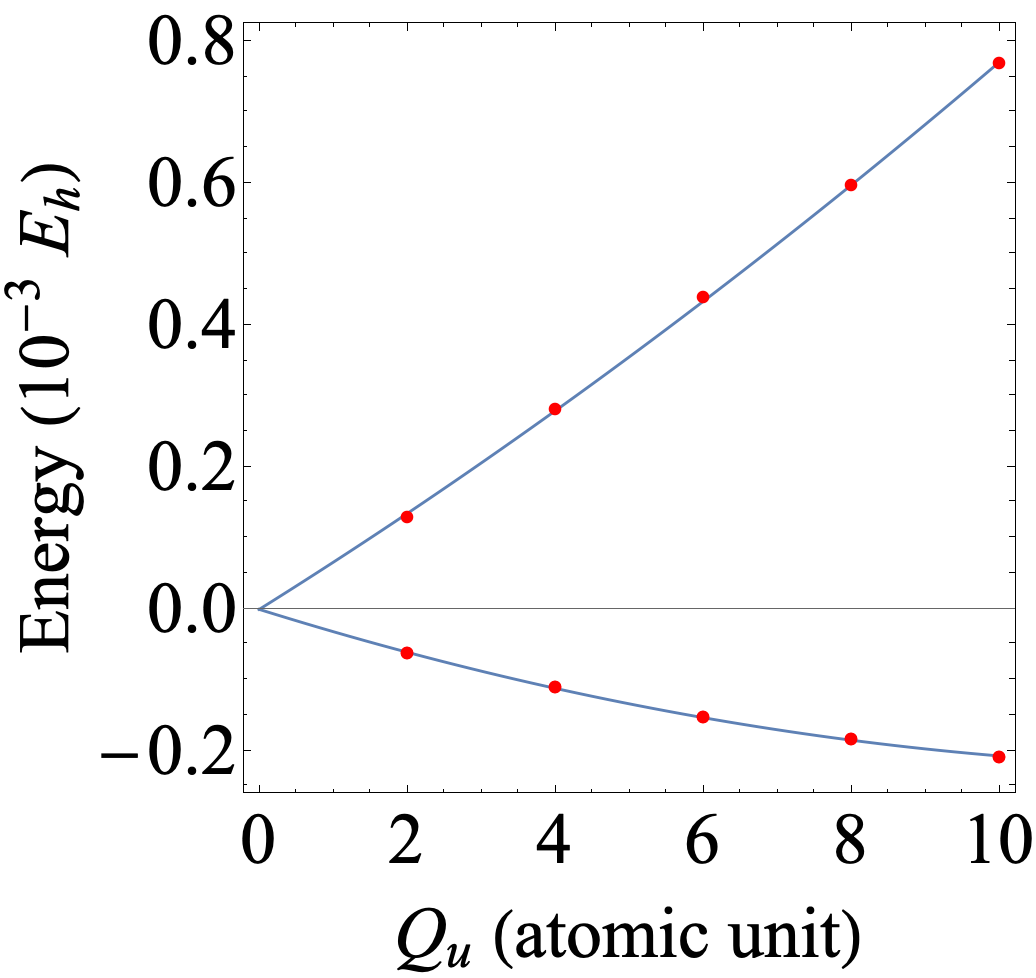}
 & &
  \raisebox{0.420\linewidth}{
 \multirow{3}{*}{
 \includegraphics[bb = 0 0 410 957, width=0.415\linewidth]{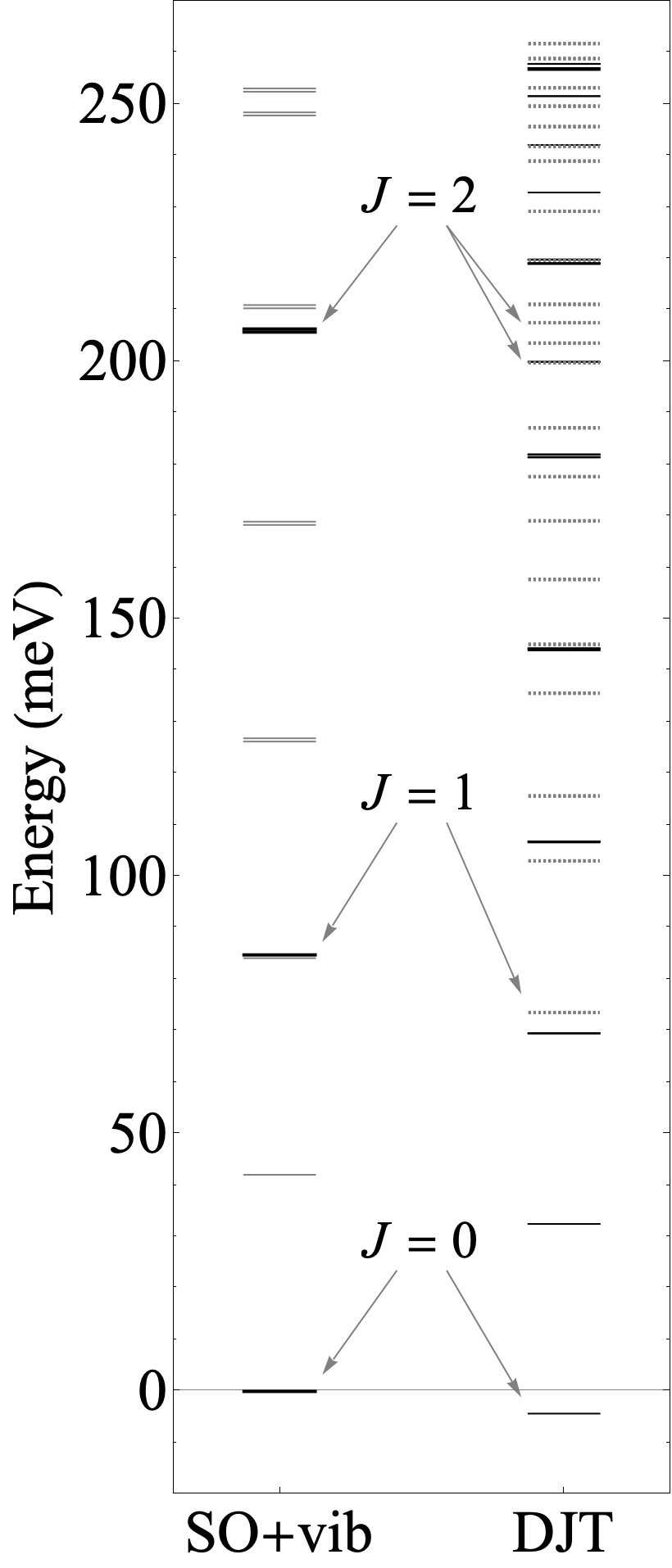}
  }
 }
\\
 (b) \\
 \includegraphics[bb = 0 0 518 470, height=0.46\linewidth]{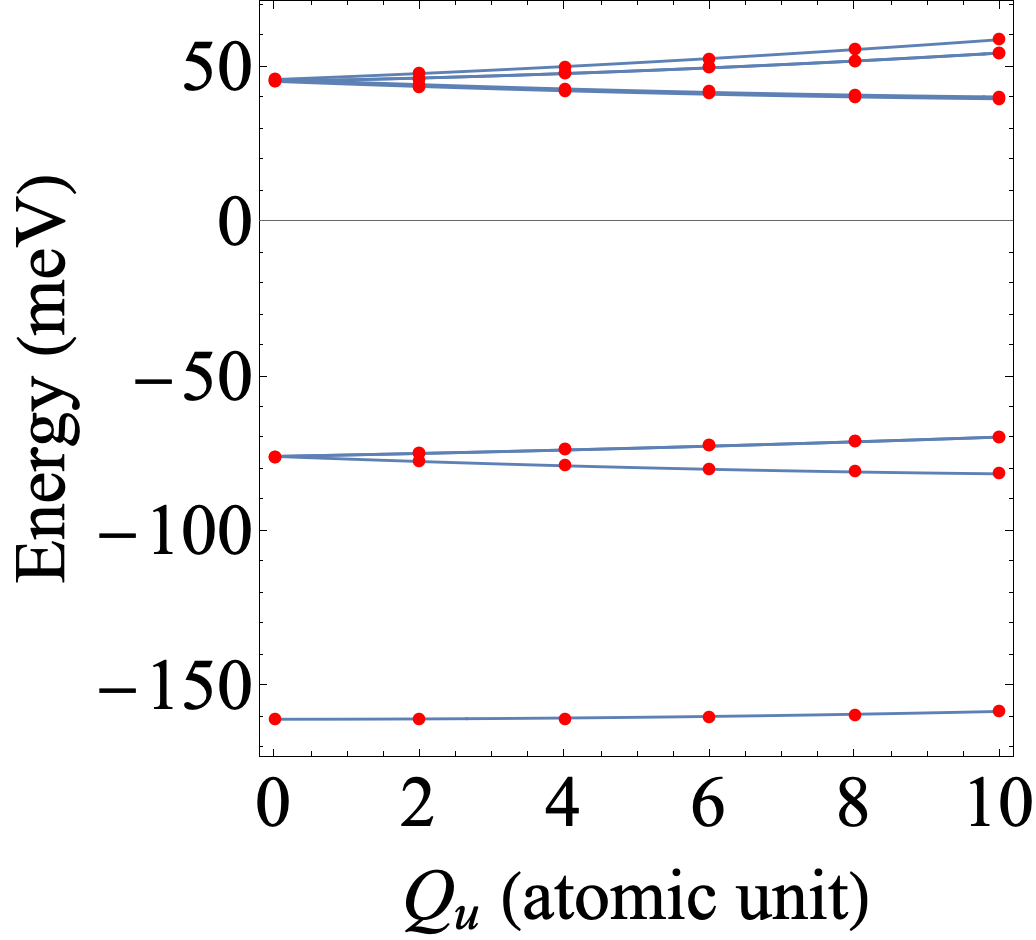}
\end{tabular}
\caption{
The adiabatic potential energies and the vibronic levels. 
(a) The $^3T_1$ term energies with respect to the JT deformation ($Q_u$) in atomic units. 
The red points are the {\it ab initio} energies and the solid lines are the eigenvalues of $U$ (\ref{Eq:U}).
(b) The spin-orbit multiplet energies (meV) with respect to the JT deformation ($Q_u$). 
(c) The spin-orbit and vibrational energy levels (SO + vib) and vibronic energy levels (DJT) (in meV). 
In the left column, the black and gray lines are the spin-orbit energies without and with vibrational excitations. 
In the right column, the solid and dashed lines are the energy eigenstates of the $(A \oplus E) \otimes E$ and the $(T_1 \oplus T_2) \otimes e$ dynamic JT models, respectively. 
}
\label{Fig:JT}
\end{figure}

\subsection{Vibronic coupling parameters}
\label{Sec:vibronic_abinitio}
We derived the vibronic coupling parameters from the gradients of the ${}^3T_1$ term energies with respect to the JT deformation. 
Figure \ref{Fig:JT}(a) indicates the {\it ab initio} ${}^3T_1$ term energies for several JT deformed structures (the red points). 
By fitting the data to Eq. (\ref{Eq:U}), we derived $\hslash \omega = 42.1$ meV and the vibronic coupling parameter $g = 1.07$.
The solid curves in Fig. \ref{Fig:JT}(a) are the best fit.

Our {\it ab initio} calculations show that the pseudo-JT couplings between the $^3T_1$ term and the other terms are weak.
We transformed the ${}^3T_1$ term states into the spin-orbit multiplet states (\ref{Eq:JM}), and draw the adiabatic potential energy surfaces in Fig. \ref{Fig:JT}(b). 
The figure indicates a good agreement between the {\it ab initio} (the red points) and the model (the blue solid lines), meaning that the pseudo JT coupling between the different multiplets is negligible.

The ${}^3T_1$ term states could vibronically couple to the $T_{2g}$ modes, while it is negligible. 
We calculated the term energies for the geometries with the $T_{2g}$ deformations, and found that the JT coupling is only a few \% of the $V$ for the $E_g$ mode. 
Therefore, we ignored the vibronic coupling to the $T_{2g}$ mode in this work.

\subsection{Vibronic states}
\label{Sec:vibronic_states_abinitio}
With the derived parameters, we calculated the vibronic states.
Figure \ref{Fig:JT}(c) shows that the vibronic coupling modulates the distribution of energy levels (the right column) with respect to the decoupled ones (the left column).
In the right column, the solid lines are the vibronic states from the $(A \oplus E) \otimes E$ JT part and the others from the $(T_1 \oplus T_2) \otimes e$ JT part. 

Now we closely look at the vibronic states which turn out to be important in the RIXS spectrum of K$_2$RuCl$_6$. 
The arrows identify the pairs of the spin-orbit and vibronic states that are close to each other.
The vibronic states have large contributions of $|\Gamma\gamma\rangle \otimes |n_u=n_v=0\rangle$ type:
the weights ($\chi_{\Gamma\gamma n_un_v, \nu}^2$) are 0.98 ($J=0$), 0.83 ($J=1$), 0.82 ($E$), and 0.78 ($T_2$).
Although the ground $J=0$ spin-orbit multiplet state does not linearly couple to the JT active vibrations, the pseudo JT coupling between the $J=0$ and $E_g$ levels (\ref{Eq:HJTJ}) stabilizes the $J=0$ vibronic state by 4 meV. 
The dynamic JT effect stabilizes the $J=1$ multiplet state by 11 meV, while it does not stabilize much the $J=2$ states due to the pseudo JT coupling between the $J=1$ and the $T_2$ part of the $J=2$ multiplet states. 

\begin{figure}[tb]
 \includegraphics[bb = 0 0 576 377, width=0.7\linewidth]{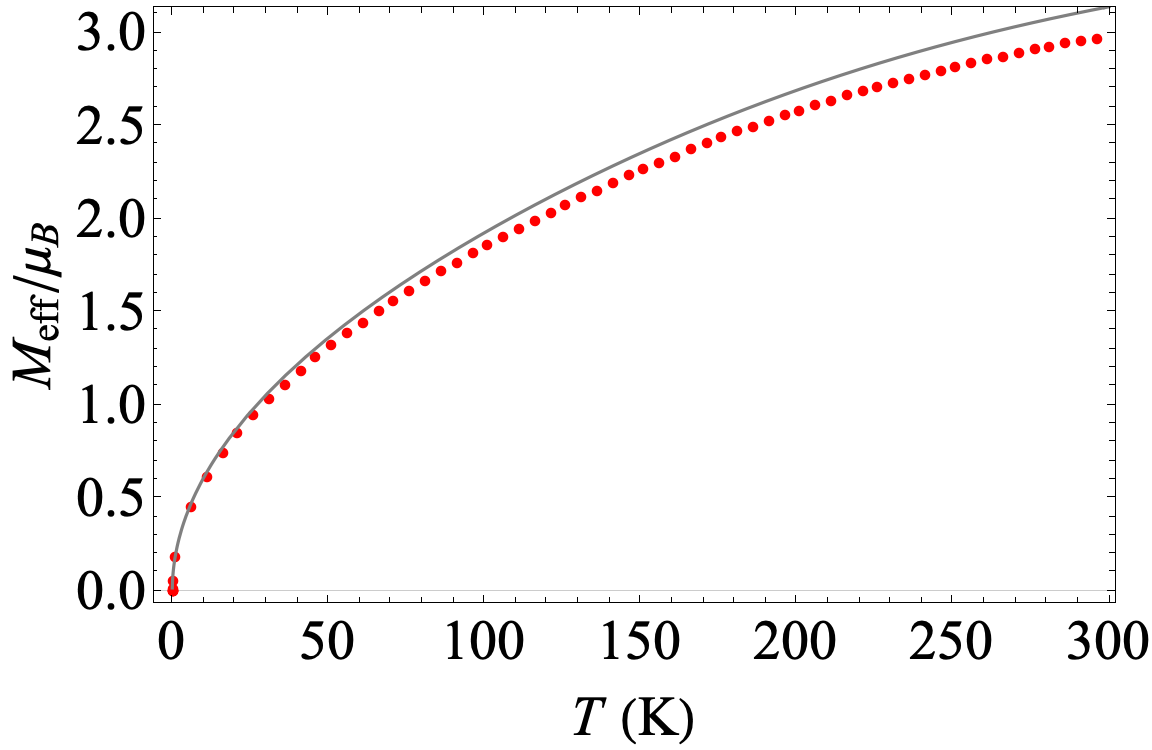}
 \caption{
 Temperature dependence of the effective magnetic moment, $M_\text{eff}$. 
 The black solid line indicates Kotani's model \cite{Kotani1949, Kotani1960} with $\lambda$ from Ref. \cite{Vishnoi2021} and the red points are the present theoretical data. 
 }
 \label{Fig:Meff}
\end{figure}

\subsection{Effective magnetic moment}
Before moving to the simulations of the RIXS spectra, let us discuss the effective magnetic moments. 
The magnetic moment operators $\hat{\mu}_\alpha$ ($\alpha = x, y, z$) within the $t_{2g}$ orbitals are 
\begin{align}
 \hat{\mu}_\alpha = \sum_{\gamma\sigma} \sum_{\gamma'\sigma'} -\mu_B \left( k \hat{l}_\alpha + g_e \hat{s}_\alpha \right)_{\gamma\sigma, \gamma'\sigma'} \hat{d}_{\gamma\sigma}^\dagger \hat{d}_{\gamma'\sigma'},
 \label{Eq:mu_orb}
\end{align}
where $\mu_B$ is the Bohr magneton, $g_e$ the g-factor of the electron, and $k$ is the reduction factor of the orbital angular momentum due to the covalency between the Ru $4d$ and Cl $3p$ orbitals. 
By fitting the {\it ab initio} magnetic moments at the CASSCF level to Eq. (\ref{Eq:mu_orb}), we determined the reduction factor $k$ to be 0.920.
Then, we projected the magnetic moments (\ref{Eq:mu_orb}) into the vibronic states (\ref{Eq:vibronic}).

With the magnetic moments, we simulated the temperature dependence of the effective magnetic moment. 
Our model consists of the vibronic Hamiltonian (\ref{Eq:H0}) and the Zeeman Hamiltonian, $\hat{H}_\text{Zee} = -\hat{\bm{\mu}} \cdot \bm{H}$, where $\bm{H}$ is the external magnetic field along the $c$ axis. 
We calculated $M_\text{eff}$ as 
\begin{align}
 M_\text{eff} &= \sqrt{3(k_BT)^2 \left.\frac{\partial^2 \text{ln}Z(\bm{H})}{\partial H^2}\right|_{H\rightarrow +0}}, 
 \label{Eq:Meff}
\end{align}
with $Z(\bm{H})$ being the partition function for the model.

Finally, we compared the calculated $M_\text{eff}$ with the experimental one from Ref. \cite{Vishnoi2021} [Fig. \ref{Fig:Meff}].
The theoretical and the experimental $M_\text{eff}$ are overall in good agreement with each other.
The deviation between them is only 5-6 \% of $M_\text{eff}$ at 300 K.
The deviation might come from the underestimations of the metal-ligand covalency ($1-k$) within the post Hartree-Fock method and Van Vleck's contribution due to the lack of the high-energy states such as $t_{2g}^3e_g^1$ within our calculations. 
The present result suggests that our model is accurate enough to adequately describe the dynamic JT effect in K$_2$RuCl$_6$.

\begin{figure}[tb]
\begin{tabular}{ll}
 (a) \\ 
 \multicolumn{2}{c}{
 \includegraphics[bb = 0 0 576 412, width=0.9\linewidth]{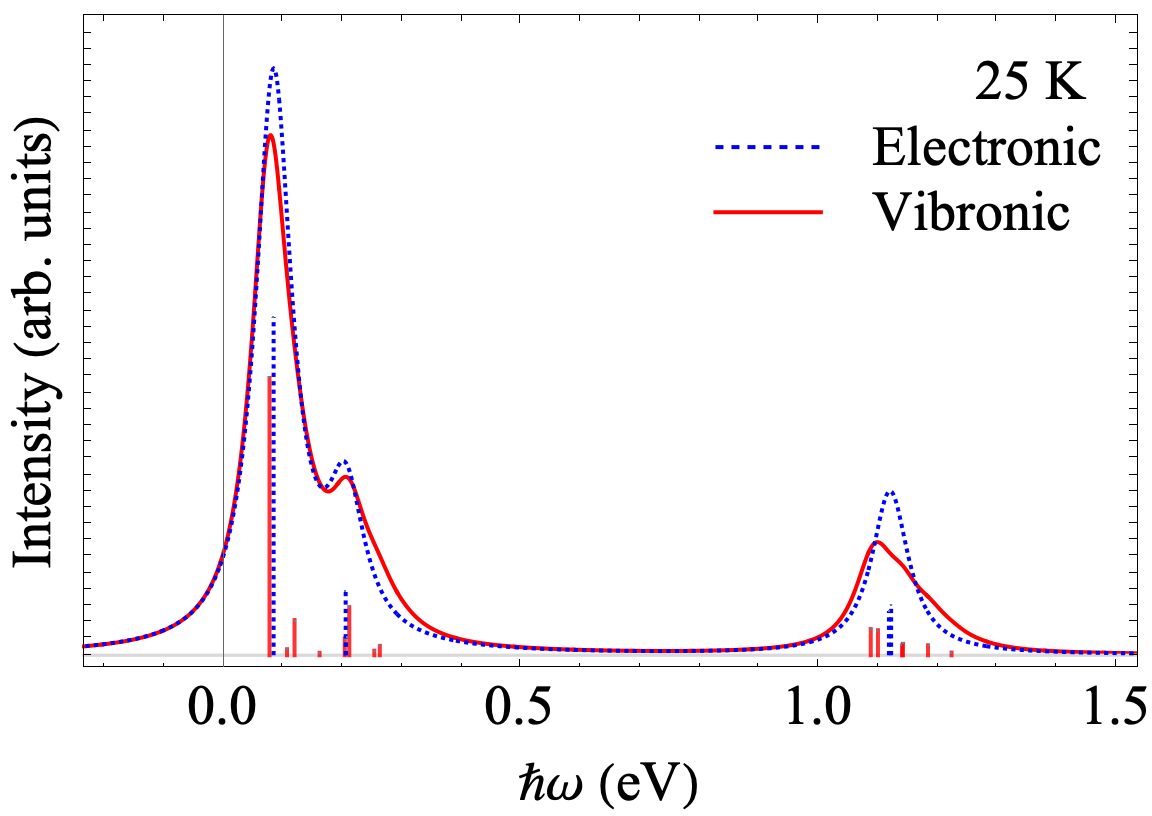}
 }
 \\
 (b) & (c) \\
\includegraphics[bb = 0 0 576 625, width=0.45\linewidth]{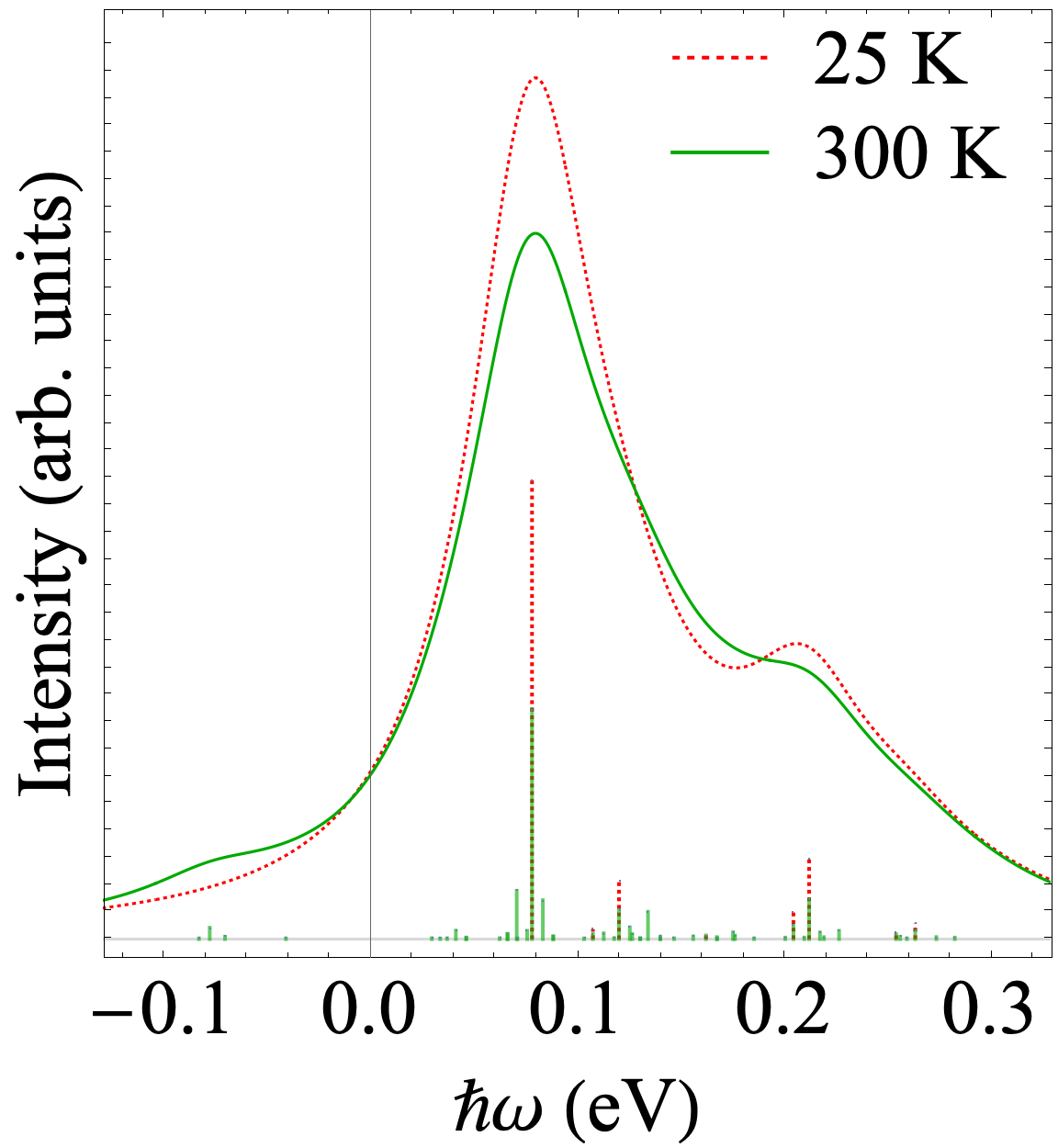}
 & 
 \includegraphics[bb = 0 0 576 621, width=0.45\linewidth]{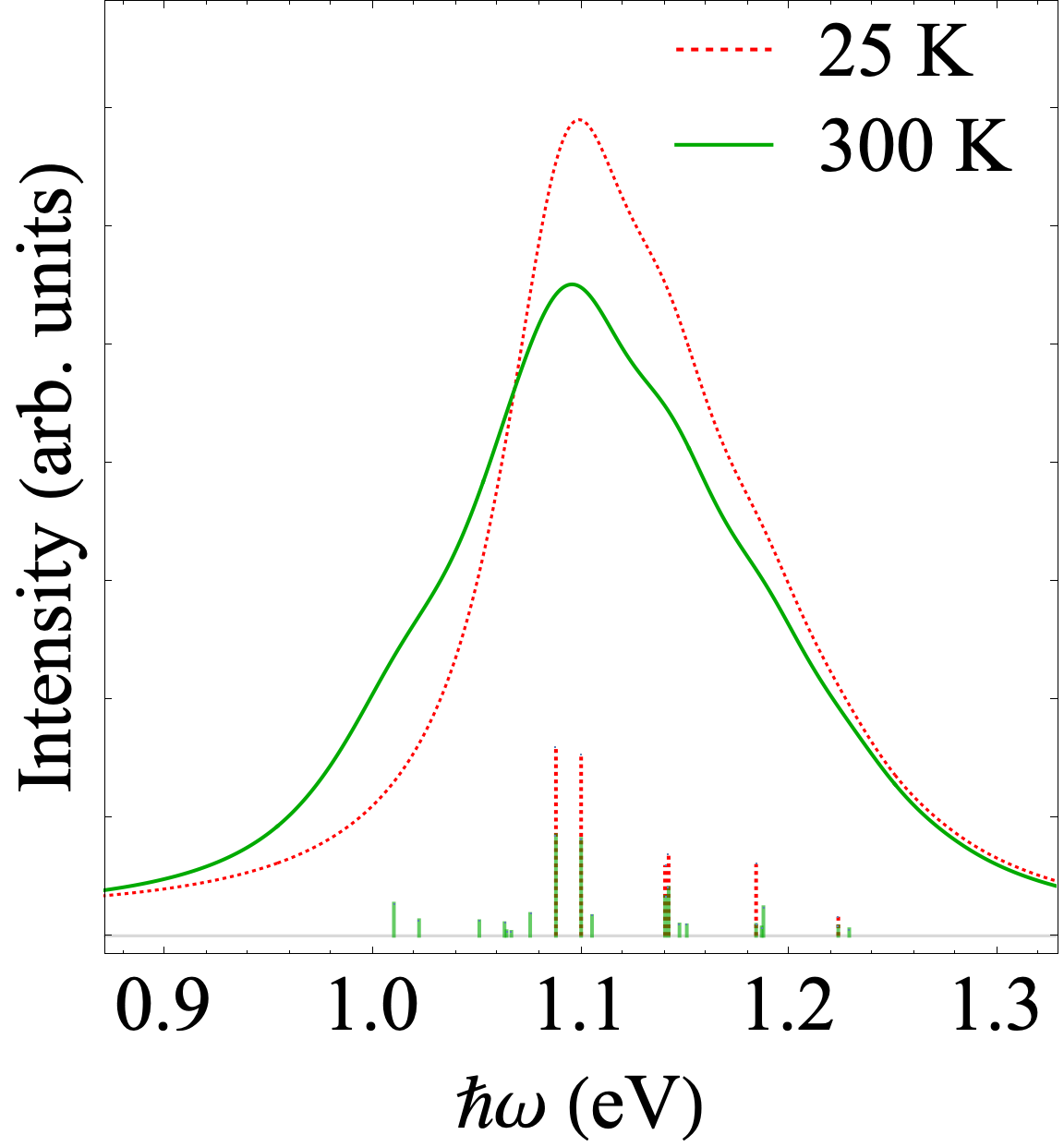}
\\
 (d)& (e) 
 \\
 \includegraphics[bb = 0 0 576 515, width=0.45\linewidth]{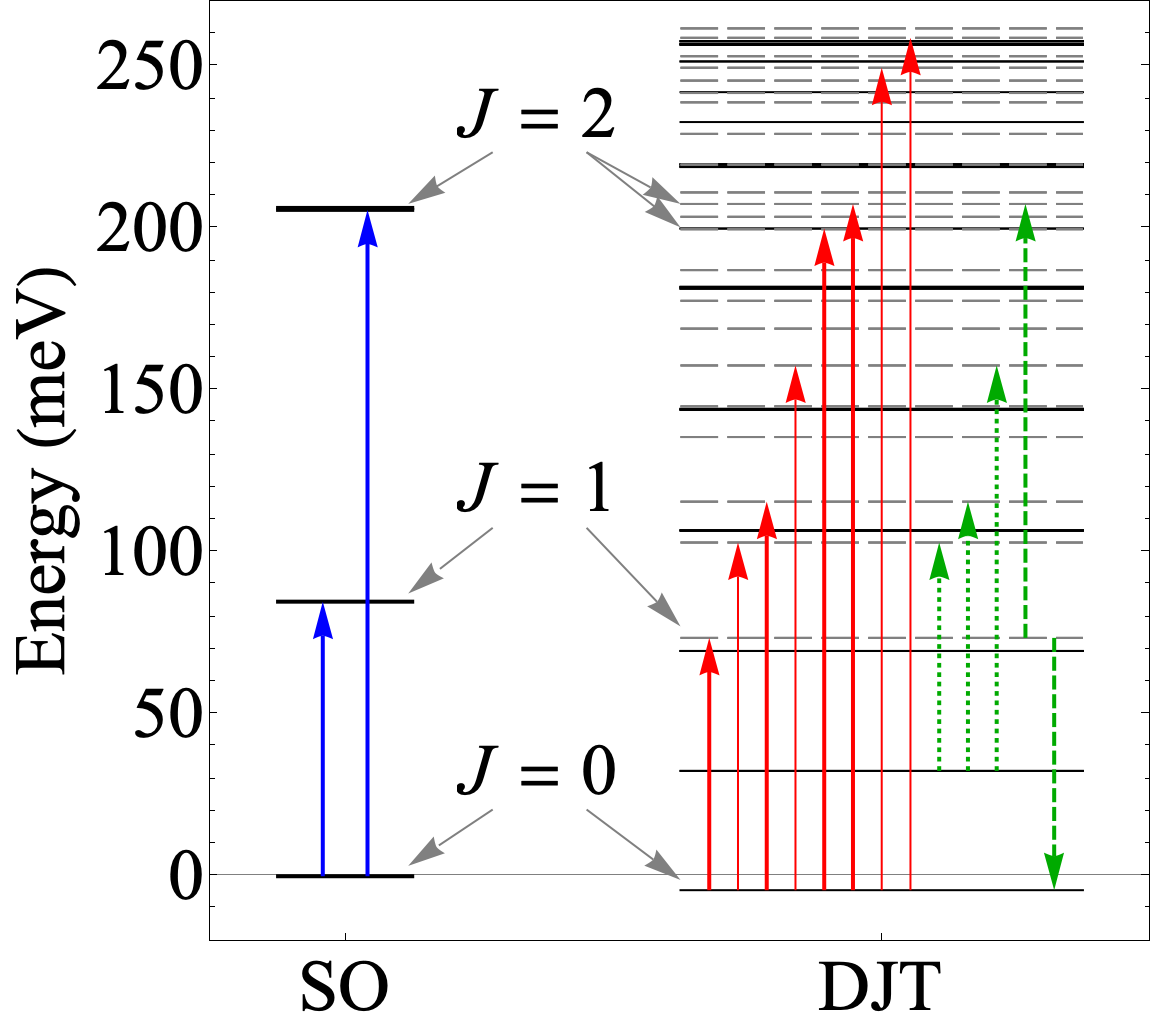}
 &
 \includegraphics[bb = 0 0 576 517, width=0.45\linewidth]{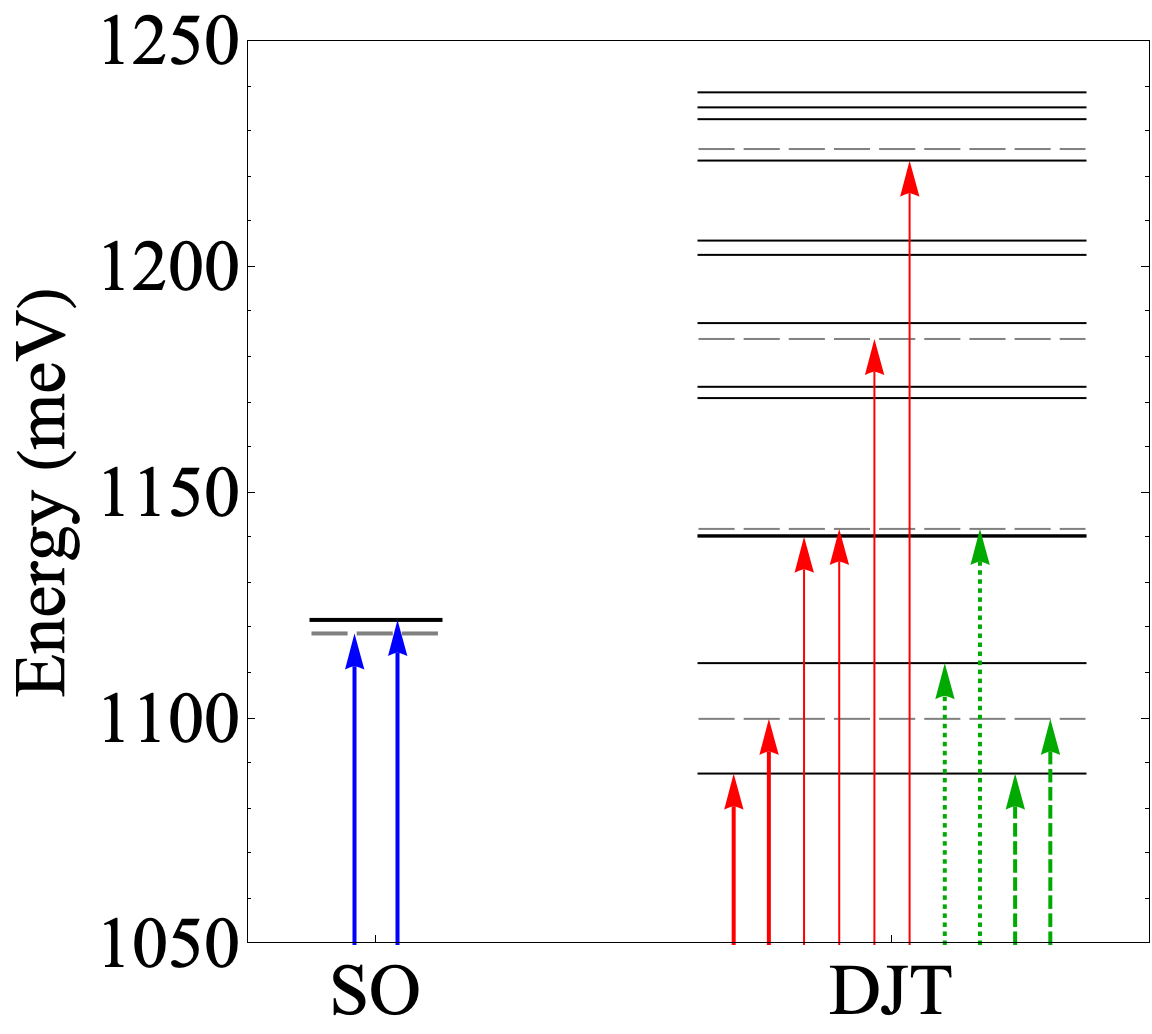}
\end{tabular}
\caption{
  RIXS spectra.
  (a) Comparison between the electronic (the blue dotted line) and vibronic (the red solid line) RIXS spectra at 25 K. 
  (b), (c) The vibronic RIXS spectra at 25 K (the red dotted line) and 300 K (the green solid line).
  $\Gamma = 0.075$ eV in all cases. 
  (d), (e) The electronic and vibronic transitions. 
}
\label{Fig:RIXS}
\end{figure}

\subsection{RIXS spectra}
\label{Sec:RIXS}
Using the numerical vibronic states in Sec. \ref{Sec:vibronic_states_abinitio}, we simulated the RIXS spectra. 
We derived the RIXS spectra by substituting the calculated vibronic states (\ref{Eq:vibronic}) into Eq. (\ref{Eq:cross_section_T}), and then convoluting the latter with Lorentzian function, $L(x) = \frac{1}{\pi} \frac{\Gamma/2}{x^2+(\Gamma/2)^2}$, where $\Gamma$ is the line width. 
For all the simulations below, $\Gamma = 0.075$ eV.
The polarizations and the directions ($\theta = \pi/4$ in the literature) of the incident and scattered lights are the same as the experimental ones \cite{Takahashi2021}.
To clarify the vibronic effect, we also calculated the RIXS spectrum with the electronic model at the same level of approximations.

Let us compare the electronic and vibronic RIXS spectra at 25 K [Fig. \ref{Fig:RIXS}(a)]. 
The strongest peak in the vibronic RIXS spectrum is lower than that in the electronic one due to the vibronic reduction of $\hat{F}$. 
The peaks in the vibronic RIXS spectrum tend to be broader than those in the electronic spectrum.
In particular, the broadening is significant at $\approx 1.1$ eV. 
We will discuss the origin below. 

The ratio of the first and second excitation energies becomes close to the experimental one due to the vibronic effect.
The peak positions of the low-energy region are at 0.078 eV and 0.212 eV, and the ratio is about 2.7, which agrees well with the experimental value \cite{Takahashi2021}. 
The ratio becomes larger than our electronic one in Sec. \ref{Sec:electronic_abinitio} because of the dynamic JT effect. 

The broadening of the RIXS spectrum occurs due to the transitions from the ground state to various excited vibronic states. 
Since the ground vibronic state $\approx |J=0\rangle \otimes |n_u=n_v=0\rangle$ in K$_2$RuCl$_6$, the main features of the peaks in the electronic RIXS spectrum persist in the vibronic RIXS spectrum, whereas more vibronic transitions exist in the latter and they make the spectra at $\approx 0.2$ eV and at $\approx 1$ eV broad. 
[Fig. \ref{Fig:RIXS}(d), (e)]. 
In particular, the peaks in the high-energy region emerge due to the presence of the dynamic JT effect rather than crystal-field splitting of the $E$ and $T_2$ multiplet levels.

As temperature rises to 300 K, the vibronic RIXS spectrum again becomes broader [Fig. \ref{Fig:RIXS}(b), (c)]. 
With the increase in temperature, the height of the peak in the low-energy region (0.078 eV) becomes lower than the one at 25 K and the peak has a new shoulder at $\approx -0.08$ eV [Fig. \ref{Fig:RIXS}(b)]. 
The peak in the high-energy region (about 1.1 eV) also becomes broader and has a new shoulder at about 1 eV [Fig. \ref{Fig:RIXS}(c)]. 
The patterns of the broadening agree well with the changes in the experimental data [Fig. \ref{Fig:k2rucl6}(d)].
The new shoulder peaks appear due to the transitions from the third excited vibronic states to the $J=0$ ground state (the green dotted line) and $E$ vibronic state (the green dashed line), respectively [Fig. \ref{Fig:RIXS}(d), (e)].

The vibronic RIXS spectrum has all the important features of the experimental spectrum, while quantitative discrepancy exists. 
The theoretical peak positions (and $\lambda$) are about 20 \% larger than the experimental data. 
The deviation comes from the underestimated covalency, and consequently overestimated $\lambda$, within the post Hartree-Fock method. 
Since all the parameters are somewhat enlarged, the qualitative features would not be affected much by the quantitative difference.

\section{Conclusion}
\label{Sec:conclusion}
We developed the {\it ab initio} based theory of RIXS spectra of a $t_{2g}^{4}$ dynamic JT ion in cubic spin-orbit Mott insulators.
We derived the electronic and vibronic parameters of an embedded Ru center by using the post Hartree-Fock calculations, and derived the low-lying vibronic states.
Using the {\it ab initio} data and Kramers-Heisenberg formula, we simulated the Ru-$L_3$ RIXS spectra.
The shape and the temperature dependence of the vibronic RIXS spectrum agree well with the experimental data, confirming the presence of the dynamic Jahn-Teller effect in K$_2$RuCl$_6$.
Our simulation indicates that several peaks emerge due to vibronic levels rather than ligand-field split spin-orbit multiplet levels. 
We also demonstrated that the dynamic JT effect enlarges the line width of the RIXS spectrum by increasing temperature.
The present results call for the reconsideration of the assignments of the RIXS spectra of cubic spin-orbit Mott insulators fully taking account of the vibronic effects.

\begin{acknowledgments}
We are grateful to Veacheslav Vieru for providing his {\it ab initio} data of the cubic cluster and reading this manuscript.
This work was partly supported by the Iketani Science and Technology Foundation and Grant-in-Aid for Scientific Research (Grant No. 22K03507) from the Japan Society for the Promotion of Science.
\end{acknowledgments}

\appendix

\section{Hole operators}
\label{A:hole}
In the hole picture, the one-electron operators $\hat{O}$ acting on the $t_{2g}$ electrons are transformed as follows:
\begin{align}
 \hat{O} &= 
 \sum_{\gamma_d\sigma}
 \sum_{\gamma_d'\sigma'}
 \langle \gamma_d\sigma| \hat{O} |\gamma_d'\sigma'\rangle 
 \hat{d}_{\gamma_d\sigma}^\dagger 
 \hat{d}_{\gamma_d'\sigma'}
 \nonumber\\
 &\rightarrow 
 \sum_{\gamma_d\sigma}
 \sum_{\gamma_d'\sigma'}
 \left[ \langle \gamma_d',\sigma'| \Theta \right] \left[ \Theta \hat{O} |\gamma_d, \sigma \rangle \right]
 \nonumber\\
 &\times
 (-1)^{s+\sigma}
 (-1)^{s+\sigma'}
 \tilde{d}_{\gamma_d-\sigma}
 \tilde{d}_{\gamma_d'-\sigma'}^\dagger 
 \nonumber\\
 &=
 -
 \sigma_\text{TR}(\hat{O})
 \sum_{\gamma_d\sigma}
 \sum_{\gamma_d'\sigma'}
 \langle \gamma_d'\sigma'| \hat{O} |\gamma_d\sigma \rangle 
 \tilde{d}_{\gamma_d'\sigma'}^\dagger 
 \tilde{d}_{\gamma_d\sigma}.
\end{align}
The time-inversion $\Theta$ of operators and spin states are \cite{Abragam1970, Sugano1970}
\begin{align}
 \Theta |\gamma_d\sigma\rangle &= (-1)^{s-\sigma} |\gamma_d, -\sigma\rangle, 
 \\
 \Theta \hat{O} &= \sigma_\text{TR}(\hat{O}) \hat{O} \Theta, 
\end{align}
respectively. 
Here $s = 1/2$, and $\sigma_\text{TR}(\hat{O})$ is the sign. 
We assumed that $\hat{O}$ is traceless. 


\begin{thebibliography}{39}%
\makeatletter
\providecommand \@ifxundefined [1]{%
 \@ifx{#1\undefined}
}%
\providecommand \@ifnum [1]{%
 \ifnum #1\expandafter \@firstoftwo
 \else \expandafter \@secondoftwo
 \fi
}%
\providecommand \@ifx [1]{%
 \ifx #1\expandafter \@firstoftwo
 \else \expandafter \@secondoftwo
 \fi
}%
\providecommand \natexlab [1]{#1}%
\providecommand \enquote  [1]{``#1''}%
\providecommand \bibnamefont  [1]{#1}%
\providecommand \bibfnamefont [1]{#1}%
\providecommand \citenamefont [1]{#1}%
\providecommand \href@noop [0]{\@secondoftwo}%
\providecommand \href [0]{\begingroup \@sanitize@url \@href}%
\providecommand \@href[1]{\@@startlink{#1}\@@href}%
\providecommand \@@href[1]{\endgroup#1\@@endlink}%
\providecommand \@sanitize@url [0]{\catcode `\\12\catcode `\$12\catcode
  `\&12\catcode `\#12\catcode `\^12\catcode `\_12\catcode `\%12\relax}%
\providecommand \@@startlink[1]{}%
\providecommand \@@endlink[0]{}%
\providecommand \url  [0]{\begingroup\@sanitize@url \@url }%
\providecommand \@url [1]{\endgroup\@href {#1}{\urlprefix }}%
\providecommand \urlprefix  [0]{URL }%
\providecommand \Eprint [0]{\href }%
\providecommand \doibase [0]{https://doi.org/}%
\providecommand \selectlanguage [0]{\@gobble}%
\providecommand \bibinfo  [0]{\@secondoftwo}%
\providecommand \bibfield  [0]{\@secondoftwo}%
\providecommand \translation [1]{[#1]}%
\providecommand \BibitemOpen [0]{}%
\providecommand \bibitemStop [0]{}%
\providecommand \bibitemNoStop [0]{.\EOS\space}%
\providecommand \EOS [0]{\spacefactor3000\relax}%
\providecommand \BibitemShut  [1]{\csname bibitem#1\endcsname}%
\let\auto@bib@innerbib\@empty
\bibitem [{\citenamefont {Witczak-Krempa}\ \emph {et~al.}(2014)\citenamefont
  {Witczak-Krempa}, \citenamefont {Chen}, \citenamefont {Kim},\ and\
  \citenamefont {Balents}}]{Witczak-Krempa2014}%
  \BibitemOpen
  \bibfield  {author} {\bibinfo {author} {\bibfnamefont {W.}~\bibnamefont
  {Witczak-Krempa}}, \bibinfo {author} {\bibfnamefont {G.}~\bibnamefont
  {Chen}}, \bibinfo {author} {\bibfnamefont {Y.~B.}\ \bibnamefont {Kim}},\ and\
  \bibinfo {author} {\bibfnamefont {L.}~\bibnamefont {Balents}},\ }\bibfield
  {title} {\bibinfo {title} {Correlated quantum phenomena in the strong
  spin-orbit regime},\ }\href
  {https://doi.org/10.1146/annurev-conmatphys-020911-125138} {\bibfield
  {journal} {\bibinfo  {journal} {Annu. Rev. Condens. Matter Phys.}\ }\textbf
  {\bibinfo {volume} {5}},\ \bibinfo {pages} {57} (\bibinfo {year}
  {2014})}\BibitemShut {NoStop}%
\bibitem [{\citenamefont {Rau}\ \emph {et~al.}(2016)\citenamefont {Rau},
  \citenamefont {Lee},\ and\ \citenamefont {Kee}}]{Rau2016}%
  \BibitemOpen
  \bibfield  {author} {\bibinfo {author} {\bibfnamefont {J.~G.}\ \bibnamefont
  {Rau}}, \bibinfo {author} {\bibfnamefont {E.~K.-H.}\ \bibnamefont {Lee}},\
  and\ \bibinfo {author} {\bibfnamefont {H.-Y.}\ \bibnamefont {Kee}},\
  }\bibfield  {title} {\bibinfo {title} {{Spin-Orbit Physics Giving Rise to
  Novel Phases in Correlated Systems: Iridates and Related Materials}},\ }\href
  {https://doi.org/10.1146/annurev-conmatphys-031115-011319} {\bibfield
  {journal} {\bibinfo  {journal} {Annu. Rev. Condens. Matter Phys.}\ }\textbf
  {\bibinfo {volume} {7}},\ \bibinfo {pages} {195} (\bibinfo {year}
  {2016})}\BibitemShut {NoStop}%
\bibitem [{\citenamefont {Takagi}\ \emph {et~al.}(2019)\citenamefont {Takagi},
  \citenamefont {Takayama}, \citenamefont {Jackeli}, \citenamefont
  {Khaliullin},\ and\ \citenamefont {Nagler}}]{Takagi2019}%
  \BibitemOpen
  \bibfield  {author} {\bibinfo {author} {\bibfnamefont {H.}~\bibnamefont
  {Takagi}}, \bibinfo {author} {\bibfnamefont {T.}~\bibnamefont {Takayama}},
  \bibinfo {author} {\bibfnamefont {G.}~\bibnamefont {Jackeli}}, \bibinfo
  {author} {\bibfnamefont {G.}~\bibnamefont {Khaliullin}},\ and\ \bibinfo
  {author} {\bibfnamefont {S.~E.}\ \bibnamefont {Nagler}},\ }\bibfield  {title}
  {\bibinfo {title} {Concept and realization of kitaev quantum spin liquids},\
  }\href {https://doi.org/10.1038/s42254-019-0038-2} {\bibfield  {journal}
  {\bibinfo  {journal} {Nat. Rev. Phys.}\ }\textbf {\bibinfo {volume} {1}},\
  \bibinfo {pages} {264} (\bibinfo {year} {2019})}\BibitemShut {NoStop}%
\bibitem [{\citenamefont {Takayama}\ \emph {et~al.}(2021)\citenamefont
  {Takayama}, \citenamefont {Chaloupka}, \citenamefont {Smerald}, \citenamefont
  {Khaliullin},\ and\ \citenamefont {Takagi}}]{Takayama2021}%
  \BibitemOpen
  \bibfield  {author} {\bibinfo {author} {\bibfnamefont {T.}~\bibnamefont
  {Takayama}}, \bibinfo {author} {\bibfnamefont {J.}~\bibnamefont {Chaloupka}},
  \bibinfo {author} {\bibfnamefont {A.}~\bibnamefont {Smerald}}, \bibinfo
  {author} {\bibfnamefont {G.}~\bibnamefont {Khaliullin}},\ and\ \bibinfo
  {author} {\bibfnamefont {H.}~\bibnamefont {Takagi}},\ }\bibfield  {title}
  {\bibinfo {title} {{Spin-Orbit-Entangled Electronic Phases in $4d$ and $5d$
  Transition-Metal Compounds}},\ }\href
  {https://doi.org/10.7566/JPSJ.90.062001} {\bibfield  {journal} {\bibinfo
  {journal} {J. Phys. Soc. Jpn.}\ }\textbf {\bibinfo {volume} {90}},\ \bibinfo
  {pages} {062001} (\bibinfo {year} {2021})}\BibitemShut {NoStop}%
\bibitem [{\citenamefont {Khaliullin}(2013)}]{Khaliullin2013}%
  \BibitemOpen
  \bibfield  {author} {\bibinfo {author} {\bibfnamefont {G.}~\bibnamefont
  {Khaliullin}},\ }\bibfield  {title} {\bibinfo {title} {Excitonic magnetism in
  van vleck--type ${d}^{4}$ mott insulators},\ }\href
  {https://doi.org/10.1103/PhysRevLett.111.197201} {\bibfield  {journal}
  {\bibinfo  {journal} {Phys. Rev. Lett.}\ }\textbf {\bibinfo {volume} {111}},\
  \bibinfo {pages} {197201} (\bibinfo {year} {2013})}\BibitemShut {NoStop}%
\bibitem [{\citenamefont {Jain}\ \emph {et~al.}(2017)\citenamefont {Jain},
  \citenamefont {Krautloher}, \citenamefont {Porras}, \citenamefont {Ryu},
  \citenamefont {Chen}, \citenamefont {Abernathy}, \citenamefont {Park},
  \citenamefont {Ivanov}, \citenamefont {Chaloupka}, \citenamefont
  {Khaliullin}, \citenamefont {Keimer},\ and\ \citenamefont {Kim}}]{Jain2017}%
  \BibitemOpen
  \bibfield  {author} {\bibinfo {author} {\bibfnamefont {A.}~\bibnamefont
  {Jain}}, \bibinfo {author} {\bibfnamefont {M.}~\bibnamefont {Krautloher}},
  \bibinfo {author} {\bibfnamefont {J.}~\bibnamefont {Porras}}, \bibinfo
  {author} {\bibfnamefont {G.~H.}\ \bibnamefont {Ryu}}, \bibinfo {author}
  {\bibfnamefont {D.~P.}\ \bibnamefont {Chen}}, \bibinfo {author}
  {\bibfnamefont {D.~L.}\ \bibnamefont {Abernathy}}, \bibinfo {author}
  {\bibfnamefont {J.~T.}\ \bibnamefont {Park}}, \bibinfo {author}
  {\bibfnamefont {A.}~\bibnamefont {Ivanov}}, \bibinfo {author} {\bibfnamefont
  {J.}~\bibnamefont {Chaloupka}}, \bibinfo {author} {\bibfnamefont
  {G.}~\bibnamefont {Khaliullin}}, \bibinfo {author} {\bibfnamefont
  {B.}~\bibnamefont {Keimer}},\ and\ \bibinfo {author} {\bibfnamefont {B.~J.}\
  \bibnamefont {Kim}},\ }\bibfield  {title} {\bibinfo {title} {Higgs mode and
  its decay in a two-dimensional antiferromagnet},\ }\href
  {https://doi.org/https://doi.org/10.1038/nphys4077} {\bibfield  {journal}
  {\bibinfo  {journal} {Nat. Phys.}\ }\textbf {\bibinfo {volume} {13}},\
  \bibinfo {pages} {633} (\bibinfo {year} {2017})}\BibitemShut {NoStop}%
\bibitem [{\citenamefont {Souliou}\ \emph {et~al.}(2017)\citenamefont
  {Souliou}, \citenamefont {Chaloupka}, \citenamefont {Khaliullin},
  \citenamefont {Ryu}, \citenamefont {Jain}, \citenamefont {Kim}, \citenamefont
  {Le~Tacon},\ and\ \citenamefont {Keimer}}]{Souliou2017}%
  \BibitemOpen
  \bibfield  {author} {\bibinfo {author} {\bibfnamefont {S.-M.}\ \bibnamefont
  {Souliou}}, \bibinfo {author} {\bibfnamefont {J.~c.~v.}\ \bibnamefont
  {Chaloupka}}, \bibinfo {author} {\bibfnamefont {G.}~\bibnamefont
  {Khaliullin}}, \bibinfo {author} {\bibfnamefont {G.}~\bibnamefont {Ryu}},
  \bibinfo {author} {\bibfnamefont {A.}~\bibnamefont {Jain}}, \bibinfo {author}
  {\bibfnamefont {B.~J.}\ \bibnamefont {Kim}}, \bibinfo {author} {\bibfnamefont
  {M.}~\bibnamefont {Le~Tacon}},\ and\ \bibinfo {author} {\bibfnamefont
  {B.}~\bibnamefont {Keimer}},\ }\bibfield  {title} {\bibinfo {title} {{Raman
  Scattering from Higgs Mode Oscillations in the Two-Dimensional
  Antiferromagnet ${\mathrm{Ca}}_{2}{\mathrm{RuO}}_{4}$}},\ }\href
  {https://doi.org/10.1103/PhysRevLett.119.067201} {\bibfield  {journal}
  {\bibinfo  {journal} {Phys. Rev. Lett.}\ }\textbf {\bibinfo {volume} {119}},\
  \bibinfo {pages} {067201} (\bibinfo {year} {2017})}\BibitemShut {NoStop}%
\bibitem [{\citenamefont {Chaloupka}\ and\ \citenamefont
  {Khaliullin}(2019)}]{Chaloupka2019}%
  \BibitemOpen
  \bibfield  {author} {\bibinfo {author} {\bibfnamefont {J.}~\bibnamefont
  {Chaloupka}}\ and\ \bibinfo {author} {\bibfnamefont {G.}~\bibnamefont
  {Khaliullin}},\ }\bibfield  {title} {\bibinfo {title} {{Highly frustrated
  magnetism in relativistic ${d}^{4}$ Mott insulators: Bosonic analog of the
  Kitaev honeycomb model}},\ }\href
  {https://doi.org/10.1103/PhysRevB.100.224413} {\bibfield  {journal} {\bibinfo
   {journal} {Phys. Rev. B}\ }\textbf {\bibinfo {volume} {100}},\ \bibinfo
  {pages} {224413} (\bibinfo {year} {2019})}\BibitemShut {NoStop}%
\bibitem [{\citenamefont {Dey}\ \emph {et~al.}(2016)\citenamefont {Dey},
  \citenamefont {Maljuk}, \citenamefont {Efremov}, \citenamefont {Kataeva},
  \citenamefont {Gass}, \citenamefont {Blum}, \citenamefont {Steckel},
  \citenamefont {Gruner}, \citenamefont {Ritschel}, \citenamefont {Wolter},
  \citenamefont {Geck}, \citenamefont {Hess}, \citenamefont {Koepernik},
  \citenamefont {van~den Brink}, \citenamefont {Wurmehl},\ and\ \citenamefont
  {B\"uchner}}]{Dey2016}%
  \BibitemOpen
  \bibfield  {author} {\bibinfo {author} {\bibfnamefont {T.}~\bibnamefont
  {Dey}}, \bibinfo {author} {\bibfnamefont {A.}~\bibnamefont {Maljuk}},
  \bibinfo {author} {\bibfnamefont {D.~V.}\ \bibnamefont {Efremov}}, \bibinfo
  {author} {\bibfnamefont {O.}~\bibnamefont {Kataeva}}, \bibinfo {author}
  {\bibfnamefont {S.}~\bibnamefont {Gass}}, \bibinfo {author} {\bibfnamefont
  {C.~G.~F.}\ \bibnamefont {Blum}}, \bibinfo {author} {\bibfnamefont
  {F.}~\bibnamefont {Steckel}}, \bibinfo {author} {\bibfnamefont
  {D.}~\bibnamefont {Gruner}}, \bibinfo {author} {\bibfnamefont
  {T.}~\bibnamefont {Ritschel}}, \bibinfo {author} {\bibfnamefont {A.~U.~B.}\
  \bibnamefont {Wolter}}, \bibinfo {author} {\bibfnamefont {J.}~\bibnamefont
  {Geck}}, \bibinfo {author} {\bibfnamefont {C.}~\bibnamefont {Hess}}, \bibinfo
  {author} {\bibfnamefont {K.}~\bibnamefont {Koepernik}}, \bibinfo {author}
  {\bibfnamefont {J.}~\bibnamefont {van~den Brink}}, \bibinfo {author}
  {\bibfnamefont {S.}~\bibnamefont {Wurmehl}},\ and\ \bibinfo {author}
  {\bibfnamefont {B.}~\bibnamefont {B\"uchner}},\ }\bibfield  {title} {\bibinfo
  {title} {{${\text{Ba}}_{2}{\text{YIrO}}_{6}$: A cubic double perovskite
  material with ${\text{Ir}}^{5+}$ ions}},\ }\href
  {https://doi.org/10.1103/PhysRevB.93.014434} {\bibfield  {journal} {\bibinfo
  {journal} {Phys. Rev. B}\ }\textbf {\bibinfo {volume} {93}},\ \bibinfo
  {pages} {014434} (\bibinfo {year} {2016})}\BibitemShut {NoStop}%
\bibitem [{\citenamefont {Yuan}\ \emph {et~al.}(2017)\citenamefont {Yuan},
  \citenamefont {Clancy}, \citenamefont {Cook}, \citenamefont {Thompson},
  \citenamefont {Greedan}, \citenamefont {Cao}, \citenamefont {Jeon},
  \citenamefont {Noh}, \citenamefont {Upton}, \citenamefont {Casa},
  \citenamefont {Gog}, \citenamefont {Paramekanti},\ and\ \citenamefont
  {Kim}}]{Yuan2017}%
  \BibitemOpen
  \bibfield  {author} {\bibinfo {author} {\bibfnamefont {B.}~\bibnamefont
  {Yuan}}, \bibinfo {author} {\bibfnamefont {J.~P.}\ \bibnamefont {Clancy}},
  \bibinfo {author} {\bibfnamefont {A.~M.}\ \bibnamefont {Cook}}, \bibinfo
  {author} {\bibfnamefont {C.~M.}\ \bibnamefont {Thompson}}, \bibinfo {author}
  {\bibfnamefont {J.}~\bibnamefont {Greedan}}, \bibinfo {author} {\bibfnamefont
  {G.}~\bibnamefont {Cao}}, \bibinfo {author} {\bibfnamefont {B.~C.}\
  \bibnamefont {Jeon}}, \bibinfo {author} {\bibfnamefont {T.~W.}\ \bibnamefont
  {Noh}}, \bibinfo {author} {\bibfnamefont {M.~H.}\ \bibnamefont {Upton}},
  \bibinfo {author} {\bibfnamefont {D.}~\bibnamefont {Casa}}, \bibinfo {author}
  {\bibfnamefont {T.}~\bibnamefont {Gog}}, \bibinfo {author} {\bibfnamefont
  {A.}~\bibnamefont {Paramekanti}},\ and\ \bibinfo {author} {\bibfnamefont
  {Y.-J.}\ \bibnamefont {Kim}},\ }\bibfield  {title} {\bibinfo {title}
  {{Determination of Hund's coupling in $5d$ oxides using resonant inelastic
  x-ray scattering}},\ }\href {https://doi.org/10.1103/PhysRevB.95.235114}
  {\bibfield  {journal} {\bibinfo  {journal} {Phys. Rev. B}\ }\textbf {\bibinfo
  {volume} {95}},\ \bibinfo {pages} {235114} (\bibinfo {year}
  {2017})}\BibitemShut {NoStop}%
\bibitem [{\citenamefont {Fuchs}\ \emph {et~al.}(2018)\citenamefont {Fuchs},
  \citenamefont {Dey}, \citenamefont {Aslan-Cansever}, \citenamefont {Maljuk},
  \citenamefont {Wurmehl}, \citenamefont {B\"uchner},\ and\ \citenamefont
  {Kataev}}]{Fuchs2018}%
  \BibitemOpen
  \bibfield  {author} {\bibinfo {author} {\bibfnamefont {S.}~\bibnamefont
  {Fuchs}}, \bibinfo {author} {\bibfnamefont {T.}~\bibnamefont {Dey}}, \bibinfo
  {author} {\bibfnamefont {G.}~\bibnamefont {Aslan-Cansever}}, \bibinfo
  {author} {\bibfnamefont {A.}~\bibnamefont {Maljuk}}, \bibinfo {author}
  {\bibfnamefont {S.}~\bibnamefont {Wurmehl}}, \bibinfo {author} {\bibfnamefont
  {B.}~\bibnamefont {B\"uchner}},\ and\ \bibinfo {author} {\bibfnamefont
  {V.}~\bibnamefont {Kataev}},\ }\bibfield  {title} {\bibinfo {title}
  {{Unraveling the Nature of Magnetism of the $5{d}^{4}$ Double Perovskite
  ${\mathrm{Ba}}_{2}{\mathrm{YIrO}}_{6}$}},\ }\href
  {https://doi.org/10.1103/PhysRevLett.120.237204} {\bibfield  {journal}
  {\bibinfo  {journal} {Phys. Rev. Lett.}\ }\textbf {\bibinfo {volume} {120}},\
  \bibinfo {pages} {237204} (\bibinfo {year} {2018})}\BibitemShut {NoStop}%
\bibitem [{\citenamefont {Takayama}\ \emph {et~al.}(2022)\citenamefont
  {Takayama}, \citenamefont {Blankenhorn}, \citenamefont {Bertinshaw},
  \citenamefont {Haskel}, \citenamefont {Bogdanov}, \citenamefont {Kitagawa},
  \citenamefont {Yaresko}, \citenamefont {Krajewska}, \citenamefont {Bette},
  \citenamefont {McNally}, \citenamefont {Gibbs}, \citenamefont {Matsumoto},
  \citenamefont {Sari}, \citenamefont {Watanabe}, \citenamefont {Fabbris},
  \citenamefont {Bi}, \citenamefont {Larkin}, \citenamefont {Rabinovich},
  \citenamefont {Boris}, \citenamefont {Ishii}, \citenamefont {Yamaoka},
  \citenamefont {Irifune}, \citenamefont {Bewley}, \citenamefont {Ridley},
  \citenamefont {Bull}, \citenamefont {Dinnebier}, \citenamefont {Keimer},\
  and\ \citenamefont {Takagi}}]{Takayama2022}%
  \BibitemOpen
  \bibfield  {author} {\bibinfo {author} {\bibfnamefont {T.}~\bibnamefont
  {Takayama}}, \bibinfo {author} {\bibfnamefont {M.}~\bibnamefont
  {Blankenhorn}}, \bibinfo {author} {\bibfnamefont {J.}~\bibnamefont
  {Bertinshaw}}, \bibinfo {author} {\bibfnamefont {D.}~\bibnamefont {Haskel}},
  \bibinfo {author} {\bibfnamefont {N.~A.}\ \bibnamefont {Bogdanov}}, \bibinfo
  {author} {\bibfnamefont {K.}~\bibnamefont {Kitagawa}}, \bibinfo {author}
  {\bibfnamefont {A.~N.}\ \bibnamefont {Yaresko}}, \bibinfo {author}
  {\bibfnamefont {A.}~\bibnamefont {Krajewska}}, \bibinfo {author}
  {\bibfnamefont {S.}~\bibnamefont {Bette}}, \bibinfo {author} {\bibfnamefont
  {G.}~\bibnamefont {McNally}}, \bibinfo {author} {\bibfnamefont {A.~S.}\
  \bibnamefont {Gibbs}}, \bibinfo {author} {\bibfnamefont {Y.}~\bibnamefont
  {Matsumoto}}, \bibinfo {author} {\bibfnamefont {D.~P.}\ \bibnamefont {Sari}},
  \bibinfo {author} {\bibfnamefont {I.}~\bibnamefont {Watanabe}}, \bibinfo
  {author} {\bibfnamefont {G.}~\bibnamefont {Fabbris}}, \bibinfo {author}
  {\bibfnamefont {W.}~\bibnamefont {Bi}}, \bibinfo {author} {\bibfnamefont
  {T.~I.}\ \bibnamefont {Larkin}}, \bibinfo {author} {\bibfnamefont {K.~S.}\
  \bibnamefont {Rabinovich}}, \bibinfo {author} {\bibfnamefont {A.~V.}\
  \bibnamefont {Boris}}, \bibinfo {author} {\bibfnamefont {H.}~\bibnamefont
  {Ishii}}, \bibinfo {author} {\bibfnamefont {H.}~\bibnamefont {Yamaoka}},
  \bibinfo {author} {\bibfnamefont {T.}~\bibnamefont {Irifune}}, \bibinfo
  {author} {\bibfnamefont {R.}~\bibnamefont {Bewley}}, \bibinfo {author}
  {\bibfnamefont {C.~J.}\ \bibnamefont {Ridley}}, \bibinfo {author}
  {\bibfnamefont {C.~L.}\ \bibnamefont {Bull}}, \bibinfo {author}
  {\bibfnamefont {R.}~\bibnamefont {Dinnebier}}, \bibinfo {author}
  {\bibfnamefont {B.}~\bibnamefont {Keimer}},\ and\ \bibinfo {author}
  {\bibfnamefont {H.}~\bibnamefont {Takagi}},\ }\bibfield  {title} {\bibinfo
  {title} {{Competing spin-orbital singlet states in the $4{d}^{4}$ honeycomb
  ruthenate ${\mathrm{Ag}}_{3}{\mathrm{LiRu}}_{2}$O$_{6}$}},\ }\href
  {https://doi.org/10.1103/PhysRevResearch.4.043079} {\bibfield  {journal}
  {\bibinfo  {journal} {Phys. Rev. Res.}\ }\textbf {\bibinfo {volume} {4}},\
  \bibinfo {pages} {043079} (\bibinfo {year} {2022})}\BibitemShut {NoStop}%
\bibitem [{\citenamefont {Takahashi}\ \emph {et~al.}(2021)\citenamefont
  {Takahashi}, \citenamefont {Suzuki}, \citenamefont {Bertinshaw},
  \citenamefont {Bette}, \citenamefont {M\"uhle}, \citenamefont {Nuss},
  \citenamefont {Dinnebier}, \citenamefont {Yaresko}, \citenamefont
  {Khaliullin}, \citenamefont {Gretarsson}, \citenamefont {Takayama},
  \citenamefont {Takagi},\ and\ \citenamefont {Keimer}}]{Takahashi2021}%
  \BibitemOpen
  \bibfield  {author} {\bibinfo {author} {\bibfnamefont {H.}~\bibnamefont
  {Takahashi}}, \bibinfo {author} {\bibfnamefont {H.}~\bibnamefont {Suzuki}},
  \bibinfo {author} {\bibfnamefont {J.}~\bibnamefont {Bertinshaw}}, \bibinfo
  {author} {\bibfnamefont {S.}~\bibnamefont {Bette}}, \bibinfo {author}
  {\bibfnamefont {C.}~\bibnamefont {M\"uhle}}, \bibinfo {author} {\bibfnamefont
  {J.}~\bibnamefont {Nuss}}, \bibinfo {author} {\bibfnamefont {R.}~\bibnamefont
  {Dinnebier}}, \bibinfo {author} {\bibfnamefont {A.}~\bibnamefont {Yaresko}},
  \bibinfo {author} {\bibfnamefont {G.}~\bibnamefont {Khaliullin}}, \bibinfo
  {author} {\bibfnamefont {H.}~\bibnamefont {Gretarsson}}, \bibinfo {author}
  {\bibfnamefont {T.}~\bibnamefont {Takayama}}, \bibinfo {author}
  {\bibfnamefont {H.}~\bibnamefont {Takagi}},\ and\ \bibinfo {author}
  {\bibfnamefont {B.}~\bibnamefont {Keimer}},\ }\bibfield  {title} {\bibinfo
  {title} {{Nonmagnetic $J=0$ State and Spin-Orbit Excitations in
  ${\mathrm{K}}_{2}{\mathrm{RuCl}}_{6}$}},\ }\href
  {https://doi.org/10.1103/PhysRevLett.127.227201} {\bibfield  {journal}
  {\bibinfo  {journal} {Phys. Rev. Lett.}\ }\textbf {\bibinfo {volume} {127}},\
  \bibinfo {pages} {227201} (\bibinfo {year} {2021})}\BibitemShut {NoStop}%
\bibitem [{\citenamefont {Hiraoka}\ \emph {et~al.}(2021)\citenamefont
  {Hiraoka}, \citenamefont {Whiteaker}, \citenamefont {Blankenhorn},
  \citenamefont {Hayashi}, \citenamefont {Oka}, \citenamefont {Takagi},\ and\
  \citenamefont {Kitagawa}}]{Hiraoka2021}%
  \BibitemOpen
  \bibfield  {author} {\bibinfo {author} {\bibfnamefont {N.}~\bibnamefont
  {Hiraoka}}, \bibinfo {author} {\bibfnamefont {K.}~\bibnamefont {Whiteaker}},
  \bibinfo {author} {\bibfnamefont {M.}~\bibnamefont {Blankenhorn}}, \bibinfo
  {author} {\bibfnamefont {Y.}~\bibnamefont {Hayashi}}, \bibinfo {author}
  {\bibfnamefont {R.}~\bibnamefont {Oka}}, \bibinfo {author} {\bibfnamefont
  {H.}~\bibnamefont {Takagi}},\ and\ \bibinfo {author} {\bibfnamefont
  {K.}~\bibnamefont {Kitagawa}},\ }\bibfield  {title} {\bibinfo {title}
  {{Design of Opposed-Anvil-Type High-Pressure Cell for Precision Magnetometry
  and Its Application to Quantum Magnetism}},\ }\href
  {https://doi.org/10.7566/JPSJ.90.074001} {\bibfield  {journal} {\bibinfo
  {journal} {J. Phys. Soc. Jpn.}\ }\textbf {\bibinfo {volume} {90}},\ \bibinfo
  {pages} {074001} (\bibinfo {year} {2021})}\BibitemShut {NoStop}%
\bibitem [{\citenamefont {Vishnoi}\ \emph {et~al.}(2021)\citenamefont
  {Vishnoi}, \citenamefont {Zuo}, \citenamefont {Cooley}, \citenamefont
  {Kautzsch}, \citenamefont {Gómez-Torres}, \citenamefont {Murillo},
  \citenamefont {Fortier}, \citenamefont {Wilson}, \citenamefont {Seshadri},\
  and\ \citenamefont {Cheetham}}]{Vishnoi2021}%
  \BibitemOpen
  \bibfield  {author} {\bibinfo {author} {\bibfnamefont {P.}~\bibnamefont
  {Vishnoi}}, \bibinfo {author} {\bibfnamefont {J.~L.}\ \bibnamefont {Zuo}},
  \bibinfo {author} {\bibfnamefont {J.~A.}\ \bibnamefont {Cooley}}, \bibinfo
  {author} {\bibfnamefont {L.}~\bibnamefont {Kautzsch}}, \bibinfo {author}
  {\bibfnamefont {A.}~\bibnamefont {Gómez-Torres}}, \bibinfo {author}
  {\bibfnamefont {J.}~\bibnamefont {Murillo}}, \bibinfo {author} {\bibfnamefont
  {S.}~\bibnamefont {Fortier}}, \bibinfo {author} {\bibfnamefont {S.~D.}\
  \bibnamefont {Wilson}}, \bibinfo {author} {\bibfnamefont {R.}~\bibnamefont
  {Seshadri}},\ and\ \bibinfo {author} {\bibfnamefont {A.~K.}\ \bibnamefont
  {Cheetham}},\ }\bibfield  {title} {\bibinfo {title} {{Chemical Control of
  Spin-Orbit Coupling and Charge Transfer in Vacancy-Ordered Ruthenium(IV)
  Halide Perovskites}},\ }\href
  {https://doi.org/https://doi.org/10.1002/anie.202013383} {\bibfield
  {journal} {\bibinfo  {journal} {Angew. Chem. Int. Ed.}\ }\textbf {\bibinfo
  {volume} {60}},\ \bibinfo {pages} {5184} (\bibinfo {year}
  {2021})}\BibitemShut {NoStop}%
\bibitem [{\citenamefont {Bersuker}\ and\ \citenamefont
  {Polinger}(1989)}]{Bersuker1989}%
  \BibitemOpen
  \bibfield  {author} {\bibinfo {author} {\bibfnamefont {I.~B.}\ \bibnamefont
  {Bersuker}}\ and\ \bibinfo {author} {\bibfnamefont {V.~Z.}\ \bibnamefont
  {Polinger}},\ }\href@noop {} {\emph {\bibinfo {title} {Vibronic Interactions
  in Molecules and Crystals}}}\ (\bibinfo  {publisher} {Springer-Verlag},\
  \bibinfo {address} {Berlin and Heidelberg},\ \bibinfo {year}
  {1989})\BibitemShut {NoStop}%
\bibitem [{\citenamefont {Bersuker}(2021)}]{Bersuker2021}%
  \BibitemOpen
  \bibfield  {author} {\bibinfo {author} {\bibfnamefont {I.~B.}\ \bibnamefont
  {Bersuker}},\ }\bibfield  {title} {\bibinfo {title} {{Jahn-Teller and
  Pseudo-Jahn-Teller Effects: From Particular Features to General Tools in
  Exploring Molecular and Solid State Properties}},\ }\href
  {https://doi.org/10.1021/acs.chemrev.0c00718} {\bibfield  {journal} {\bibinfo
   {journal} {Chem. Rev.}\ }\textbf {\bibinfo {volume} {121}},\ \bibinfo
  {pages} {1463} (\bibinfo {year} {2021})}\BibitemShut {NoStop}%
\bibitem [{\citenamefont {Liu}\ and\ \citenamefont
  {Khaliullin}(2019)}]{Liu2019}%
  \BibitemOpen
  \bibfield  {author} {\bibinfo {author} {\bibfnamefont {H.}~\bibnamefont
  {Liu}}\ and\ \bibinfo {author} {\bibfnamefont {G.}~\bibnamefont
  {Khaliullin}},\ }\bibfield  {title} {\bibinfo {title} {{Pseudo-Jahn-Teller
  Effect and Magnetoelastic Coupling in Spin-Orbit Mott Insulators}},\ }\href
  {https://doi.org/10.1103/PhysRevLett.122.057203} {\bibfield  {journal}
  {\bibinfo  {journal} {Phys. Rev. Lett.}\ }\textbf {\bibinfo {volume} {122}},\
  \bibinfo {pages} {057203} (\bibinfo {year} {2019})}\BibitemShut {NoStop}%
\bibitem [{\citenamefont {Suzuki}\ \emph {et~al.}(2021)\citenamefont {Suzuki},
  \citenamefont {Liu}, \citenamefont {Bertinshaw}, \citenamefont {Ueda},
  \citenamefont {Kim}, \citenamefont {Laha}, \citenamefont {Weber},
  \citenamefont {Yang}, \citenamefont {Wang}, \citenamefont {Takahashi},
  \citenamefont {F\"{u}rsich}, \citenamefont {Minola}, \citenamefont {Lotsch},
  \citenamefont {Kim}, \citenamefont {Yavaş}, \citenamefont {Daghofer},
  \citenamefont {Chaloupka}, \citenamefont {Khaliullin}, \citenamefont
  {Gretarsson},\ and\ \citenamefont {Keimer}}]{Suzuki2021}%
  \BibitemOpen
  \bibfield  {author} {\bibinfo {author} {\bibfnamefont {H.}~\bibnamefont
  {Suzuki}}, \bibinfo {author} {\bibfnamefont {H.}~\bibnamefont {Liu}},
  \bibinfo {author} {\bibfnamefont {J.}~\bibnamefont {Bertinshaw}}, \bibinfo
  {author} {\bibfnamefont {K.}~\bibnamefont {Ueda}}, \bibinfo {author}
  {\bibfnamefont {H.}~\bibnamefont {Kim}}, \bibinfo {author} {\bibfnamefont
  {S.}~\bibnamefont {Laha}}, \bibinfo {author} {\bibfnamefont {D.}~\bibnamefont
  {Weber}}, \bibinfo {author} {\bibfnamefont {Z.}~\bibnamefont {Yang}},
  \bibinfo {author} {\bibfnamefont {L.}~\bibnamefont {Wang}}, \bibinfo {author}
  {\bibfnamefont {H.}~\bibnamefont {Takahashi}}, \bibinfo {author}
  {\bibfnamefont {K.}~\bibnamefont {F\"{u}rsich}}, \bibinfo {author}
  {\bibfnamefont {M.}~\bibnamefont {Minola}}, \bibinfo {author} {\bibfnamefont
  {B.~V.}\ \bibnamefont {Lotsch}}, \bibinfo {author} {\bibfnamefont {B.~J.}\
  \bibnamefont {Kim}}, \bibinfo {author} {\bibfnamefont {H.}~\bibnamefont
  {Yavaş}}, \bibinfo {author} {\bibfnamefont {M.}~\bibnamefont {Daghofer}},
  \bibinfo {author} {\bibfnamefont {J.}~\bibnamefont {Chaloupka}}, \bibinfo
  {author} {\bibfnamefont {G.}~\bibnamefont {Khaliullin}}, \bibinfo {author}
  {\bibfnamefont {H.}~\bibnamefont {Gretarsson}},\ and\ \bibinfo {author}
  {\bibfnamefont {B.}~\bibnamefont {Keimer}},\ }\bibfield  {title} {\bibinfo
  {title} {{Proximate ferromagnetic state in the Kitaev model material
  $\alpha$-RuCl$_3$}},\ }\href {https://doi.org/10.1038/s41467-021-24722-4}
  {\bibfield  {journal} {\bibinfo  {journal} {Nat. Commun.}\ }\textbf {\bibinfo
  {volume} {12}},\ \bibinfo {pages} {4512} (\bibinfo {year}
  {2021})}\BibitemShut {NoStop}%
\bibitem [{\citenamefont {Sugano}\ \emph {et~al.}(1970)\citenamefont {Sugano},
  \citenamefont {Tanabe},\ and\ \citenamefont {Kamimura}}]{Sugano1970}%
  \BibitemOpen
  \bibfield  {author} {\bibinfo {author} {\bibfnamefont {S.}~\bibnamefont
  {Sugano}}, \bibinfo {author} {\bibfnamefont {Y.}~\bibnamefont {Tanabe}},\
  and\ \bibinfo {author} {\bibfnamefont {H.}~\bibnamefont {Kamimura}},\
  }\href@noop {} {\emph {\bibinfo {title} {{Multiplets of Transition-Metal Ions
  in Crystals}}}}\ (\bibinfo  {publisher} {Academic Press},\ \bibinfo {address}
  {New York},\ \bibinfo {year} {1970})\BibitemShut {NoStop}%
\bibitem [{\citenamefont {Englman}(1972)}]{Englman1972}%
  \BibitemOpen
  \bibfield  {author} {\bibinfo {author} {\bibfnamefont {R.}~\bibnamefont
  {Englman}},\ }\href@noop {} {\emph {\bibinfo {title} {The Jahn-Teller Effect
  in Molecules and Crystals}}}\ (\bibinfo  {publisher} {John Wiley \& Sons
  Ltd},\ \bibinfo {address} {London},\ \bibinfo {year} {1972})\BibitemShut
  {NoStop}%
\bibitem [{\citenamefont {Inui}\ \emph {et~al.}(1990)\citenamefont {Inui},
  \citenamefont {Tanabe},\ and\ \citenamefont {Onodera}}]{Inui1990}%
  \BibitemOpen
  \bibfield  {author} {\bibinfo {author} {\bibfnamefont {T.}~\bibnamefont
  {Inui}}, \bibinfo {author} {\bibfnamefont {Y.}~\bibnamefont {Tanabe}},\ and\
  \bibinfo {author} {\bibfnamefont {Y.}~\bibnamefont {Onodera}},\ }\href@noop
  {} {\emph {\bibinfo {title} {Group Theory and Its Applications in Physics}}}\
  (\bibinfo  {publisher} {Springer-Verlag},\ \bibinfo {address} {Berlin and
  Heidelberg},\ \bibinfo {year} {1990})\BibitemShut {NoStop}%
\bibitem [{\citenamefont {Koster}\ \emph {et~al.}(1963)\citenamefont {Koster},
  \citenamefont {Dimmock}, \citenamefont {Wheeler},\ and\ \citenamefont
  {Statz}}]{Koster1963}%
  \BibitemOpen
  \bibfield  {author} {\bibinfo {author} {\bibfnamefont {G.~F.}\ \bibnamefont
  {Koster}}, \bibinfo {author} {\bibfnamefont {J.~O.}\ \bibnamefont {Dimmock}},
  \bibinfo {author} {\bibfnamefont {R.~G.}\ \bibnamefont {Wheeler}},\ and\
  \bibinfo {author} {\bibfnamefont {H.}~\bibnamefont {Statz}},\ }\href@noop {}
  {\emph {\bibinfo {title} {Properties of the thirty-two point groups}}}\
  (\bibinfo  {publisher} {MIT press},\ \bibinfo {address} {Massachusetts},\
  \bibinfo {year} {1963})\BibitemShut {NoStop}%
\bibitem [{\citenamefont {Varshalovich}\ \emph {et~al.}(1988)\citenamefont
  {Varshalovich}, \citenamefont {Moskalev},\ and\ \citenamefont
  {Khersonskii}}]{Varshalovich1988}%
  \BibitemOpen
  \bibfield  {author} {\bibinfo {author} {\bibfnamefont {D.~A.}\ \bibnamefont
  {Varshalovich}}, \bibinfo {author} {\bibfnamefont {A.~N.}\ \bibnamefont
  {Moskalev}},\ and\ \bibinfo {author} {\bibfnamefont {V.~K.}\ \bibnamefont
  {Khersonskii}},\ }\href@noop {} {\emph {\bibinfo {title} {{Quantum Theory of
  Angular Momentum}}}}\ (\bibinfo  {publisher} {World Scientific},\ \bibinfo
  {address} {Singapore},\ \bibinfo {year} {1988})\BibitemShut {NoStop}%
\bibitem [{\citenamefont {Sakurai}(1967)}]{Sakurai1967}%
  \BibitemOpen
  \bibfield  {author} {\bibinfo {author} {\bibfnamefont {J.~J.}\ \bibnamefont
  {Sakurai}},\ }\href@noop {} {\emph {\bibinfo {title} {Advanced Quantum
  Mechanics}}}\ (\bibinfo  {publisher} {Addison-Wesley},\ \bibinfo {address}
  {Massachusetts},\ \bibinfo {year} {1967})\BibitemShut {NoStop}%
\bibitem [{\citenamefont {Luo}\ \emph {et~al.}(1993)\citenamefont {Luo},
  \citenamefont {Trammell},\ and\ \citenamefont {Hannon}}]{Luo1993}%
  \BibitemOpen
  \bibfield  {author} {\bibinfo {author} {\bibfnamefont {J.}~\bibnamefont
  {Luo}}, \bibinfo {author} {\bibfnamefont {G.~T.}\ \bibnamefont {Trammell}},\
  and\ \bibinfo {author} {\bibfnamefont {J.~P.}\ \bibnamefont {Hannon}},\
  }\bibfield  {title} {\bibinfo {title} {Scattering operator for elastic and
  inelastic resonant x-ray scattering},\ }\href
  {https://doi.org/10.1103/PhysRevLett.71.287} {\bibfield  {journal} {\bibinfo
  {journal} {Phys. Rev. Lett.}\ }\textbf {\bibinfo {volume} {71}},\ \bibinfo
  {pages} {287} (\bibinfo {year} {1993})}\BibitemShut {NoStop}%
\bibitem [{\citenamefont {van Veenendaal}(2006)}]{vanVeenendaal2006}%
  \BibitemOpen
  \bibfield  {author} {\bibinfo {author} {\bibfnamefont {M.}~\bibnamefont {van
  Veenendaal}},\ }\bibfield  {title} {\bibinfo {title} {{Polarization
  Dependence of $L$- and $M$-Edge Resonant Inelastic X-Ray Scattering in
  Transition-Metal Compounds}},\ }\href
  {https://doi.org/10.1103/PhysRevLett.96.117404} {\bibfield  {journal}
  {\bibinfo  {journal} {Phys. Rev. Lett.}\ }\textbf {\bibinfo {volume} {96}},\
  \bibinfo {pages} {117404} (\bibinfo {year} {2006})}\BibitemShut {NoStop}%
\bibitem [{\citenamefont {Seijo}\ and\ \citenamefont
  {Barandiar{\'a}n}(1999)}]{Seijo1999}%
  \BibitemOpen
  \bibfield  {author} {\bibinfo {author} {\bibfnamefont {L.}~\bibnamefont
  {Seijo}}\ and\ \bibinfo {author} {\bibfnamefont {Z.}~\bibnamefont
  {Barandiar{\'a}n}},\ }\bibinfo {title} {{Computational Modelling of the
  Magnetic Properties of Lanthanide Compounds}},\ in\ \href@noop {} {\emph
  {\bibinfo {booktitle} {Computational Chemistry: Reviews of Current
  Trends}}},\ Vol.~\bibinfo {volume} {4},\ \bibinfo {editor} {edited by\
  \bibinfo {editor} {\bibfnamefont {J.}~\bibnamefont {Leszczynski}}}\ (\bibinfo
   {publisher} {World Scientific},\ \bibinfo {address} {Singapore},\ \bibinfo
  {year} {1999})\ pp.\ \bibinfo {pages} {55--152}\BibitemShut {NoStop}%
\bibitem [{\citenamefont {Roos}\ \emph {et~al.}(2016)\citenamefont {Roos},
  \citenamefont {Lindh}, \citenamefont {Malmqvist}, \citenamefont {Veryazov},\
  and\ \citenamefont {Widmark}}]{Roos2016}%
  \BibitemOpen
  \bibfield  {author} {\bibinfo {author} {\bibfnamefont {B.~O.}\ \bibnamefont
  {Roos}}, \bibinfo {author} {\bibfnamefont {R.}~\bibnamefont {Lindh}},
  \bibinfo {author} {\bibfnamefont {P.-{\AA}.}\ \bibnamefont {Malmqvist}},
  \bibinfo {author} {\bibfnamefont {V.}~\bibnamefont {Veryazov}},\ and\
  \bibinfo {author} {\bibfnamefont {P.-O.}\ \bibnamefont {Widmark}},\
  }\href@noop {} {\emph {\bibinfo {title} {Multiconfigurational Quantum
  Chemistry}}}\ (\bibinfo  {publisher} {Wiley},\ \bibinfo {address} {New
  Jersey},\ \bibinfo {year} {2016})\BibitemShut {NoStop}%
\bibitem [{\citenamefont {Granovsky}(2011)}]{Granovsky2011}%
  \BibitemOpen
  \bibfield  {author} {\bibinfo {author} {\bibfnamefont {A.~A.}\ \bibnamefont
  {Granovsky}},\ }\bibfield  {title} {\bibinfo {title} {Extended
  multi-configuration quasi-degenerate perturbation theory: The new approach to
  multi-state multi-reference perturbation theory},\ }\href
  {http://scitation.aip.org/content/aip/journal/jcp/134/21/10.1063/1.3596699}
  {\bibfield  {journal} {\bibinfo  {journal} {J. Chem. Phys.}\ }\textbf
  {\bibinfo {volume} {134}},\ \bibinfo {eid} {214113} (\bibinfo {year}
  {2011})}\BibitemShut {NoStop}%
\bibitem [{\citenamefont {Shiozaki}\ \emph {et~al.}(2011)\citenamefont
  {Shiozaki}, \citenamefont {Gy\H{o}rffy}, \citenamefont {Celani},\ and\
  \citenamefont {Werner}}]{Shiozaki2011}%
  \BibitemOpen
  \bibfield  {author} {\bibinfo {author} {\bibfnamefont {T.}~\bibnamefont
  {Shiozaki}}, \bibinfo {author} {\bibfnamefont {W.}~\bibnamefont
  {Gy\H{o}rffy}}, \bibinfo {author} {\bibfnamefont {P.}~\bibnamefont
  {Celani}},\ and\ \bibinfo {author} {\bibfnamefont {H.-J.}\ \bibnamefont
  {Werner}},\ }\bibfield  {title} {\bibinfo {title} {Extended multi-state
  complete active space second-order perturbation theory: Energy and nuclear
  gradients},\ }\href
  {http://scitation.aip.org/content/aip/journal/jcp/135/8/10.1063/1.3633329}
  {\bibfield  {journal} {\bibinfo  {journal} {J. Chem. Phys.}\ }\textbf
  {\bibinfo {volume} {135}},\ \bibinfo {eid} {081106} (\bibinfo {year}
  {2011})}\BibitemShut {NoStop}%
\bibitem [{\citenamefont {Fdez.~Galv\'{a}n}\ \emph {et~al.}(2019)\citenamefont
  {Fdez.~Galv\'{a}n}, \citenamefont {Vacher}, \citenamefont {Alavi},
  \citenamefont {Angeli}, \citenamefont {Aquilante}, \citenamefont
  {Autschbach}, \citenamefont {Bao}, \citenamefont {Bokarev}, \citenamefont
  {Bogdanov}, \citenamefont {Carlson}, \citenamefont {Chibotaru}, \citenamefont
  {Creutzberg}, \citenamefont {Dattani}, \citenamefont {Delcey}, \citenamefont
  {Dong}, \citenamefont {Dreuw}, \citenamefont {Freitag}, \citenamefont
  {Frutos}, \citenamefont {Gagliardi}, \citenamefont {Gendron}, \citenamefont
  {Giussani}, \citenamefont {González}, \citenamefont {Grell}, \citenamefont
  {Guo}, \citenamefont {Hoyer}, \citenamefont {Johansson}, \citenamefont
  {Keller}, \citenamefont {Knecht}, \citenamefont {Kovačević}, \citenamefont
  {Källman}, \citenamefont {Li~Manni}, \citenamefont {Lundberg}, \citenamefont
  {Ma}, \citenamefont {Mai}, \citenamefont {Malhado}, \citenamefont
  {Malmqvist}, \citenamefont {Marquetand}, \citenamefont {Mewes}, \citenamefont
  {Norell}, \citenamefont {Olivucci}, \citenamefont {Oppel}, \citenamefont
  {Phung}, \citenamefont {Pierloot}, \citenamefont {Plasser}, \citenamefont
  {Reiher}, \citenamefont {Sand}, \citenamefont {Schapiro}, \citenamefont
  {Sharma}, \citenamefont {Stein}, \citenamefont {Sørensen}, \citenamefont
  {Truhlar}, \citenamefont {Ugandi}, \citenamefont {Ungur}, \citenamefont
  {Valentini}, \citenamefont {Vancoillie}, \citenamefont {Veryazov},
  \citenamefont {Weser}, \citenamefont {Wesołowski}, \citenamefont {Widmark},
  \citenamefont {Wouters}, \citenamefont {Zech}, \citenamefont {Zobel},\ and\
  \citenamefont {Lindh}}]{molcas1}%
  \BibitemOpen
  \bibfield  {author} {\bibinfo {author} {\bibfnamefont {I.}~\bibnamefont
  {Fdez.~Galv\'{a}n}}, \bibinfo {author} {\bibfnamefont {M.}~\bibnamefont
  {Vacher}}, \bibinfo {author} {\bibfnamefont {A.}~\bibnamefont {Alavi}},
  \bibinfo {author} {\bibfnamefont {C.}~\bibnamefont {Angeli}}, \bibinfo
  {author} {\bibfnamefont {F.}~\bibnamefont {Aquilante}}, \bibinfo {author}
  {\bibfnamefont {J.}~\bibnamefont {Autschbach}}, \bibinfo {author}
  {\bibfnamefont {J.~J.}\ \bibnamefont {Bao}}, \bibinfo {author} {\bibfnamefont
  {S.~I.}\ \bibnamefont {Bokarev}}, \bibinfo {author} {\bibfnamefont {N.~A.}\
  \bibnamefont {Bogdanov}}, \bibinfo {author} {\bibfnamefont {R.~K.}\
  \bibnamefont {Carlson}}, \bibinfo {author} {\bibfnamefont {L.~F.}\
  \bibnamefont {Chibotaru}}, \bibinfo {author} {\bibfnamefont {J.}~\bibnamefont
  {Creutzberg}}, \bibinfo {author} {\bibfnamefont {N.}~\bibnamefont {Dattani}},
  \bibinfo {author} {\bibfnamefont {M.~G.}\ \bibnamefont {Delcey}}, \bibinfo
  {author} {\bibfnamefont {S.~S.}\ \bibnamefont {Dong}}, \bibinfo {author}
  {\bibfnamefont {A.}~\bibnamefont {Dreuw}}, \bibinfo {author} {\bibfnamefont
  {L.}~\bibnamefont {Freitag}}, \bibinfo {author} {\bibfnamefont {L.~M.}\
  \bibnamefont {Frutos}}, \bibinfo {author} {\bibfnamefont {L.}~\bibnamefont
  {Gagliardi}}, \bibinfo {author} {\bibfnamefont {F.}~\bibnamefont {Gendron}},
  \bibinfo {author} {\bibfnamefont {A.}~\bibnamefont {Giussani}}, \bibinfo
  {author} {\bibfnamefont {L.}~\bibnamefont {González}}, \bibinfo {author}
  {\bibfnamefont {G.}~\bibnamefont {Grell}}, \bibinfo {author} {\bibfnamefont
  {M.}~\bibnamefont {Guo}}, \bibinfo {author} {\bibfnamefont {C.~E.}\
  \bibnamefont {Hoyer}}, \bibinfo {author} {\bibfnamefont {M.}~\bibnamefont
  {Johansson}}, \bibinfo {author} {\bibfnamefont {S.}~\bibnamefont {Keller}},
  \bibinfo {author} {\bibfnamefont {S.}~\bibnamefont {Knecht}}, \bibinfo
  {author} {\bibfnamefont {G.}~\bibnamefont {Kovačević}}, \bibinfo {author}
  {\bibfnamefont {E.}~\bibnamefont {Källman}}, \bibinfo {author}
  {\bibfnamefont {G.}~\bibnamefont {Li~Manni}}, \bibinfo {author}
  {\bibfnamefont {M.}~\bibnamefont {Lundberg}}, \bibinfo {author}
  {\bibfnamefont {Y.}~\bibnamefont {Ma}}, \bibinfo {author} {\bibfnamefont
  {S.}~\bibnamefont {Mai}}, \bibinfo {author} {\bibfnamefont {J.~P.}\
  \bibnamefont {Malhado}}, \bibinfo {author} {\bibfnamefont {P.~{\AA}.}\
  \bibnamefont {Malmqvist}}, \bibinfo {author} {\bibfnamefont {P.}~\bibnamefont
  {Marquetand}}, \bibinfo {author} {\bibfnamefont {S.~A.}\ \bibnamefont
  {Mewes}}, \bibinfo {author} {\bibfnamefont {J.}~\bibnamefont {Norell}},
  \bibinfo {author} {\bibfnamefont {M.}~\bibnamefont {Olivucci}}, \bibinfo
  {author} {\bibfnamefont {M.}~\bibnamefont {Oppel}}, \bibinfo {author}
  {\bibfnamefont {Q.~M.}\ \bibnamefont {Phung}}, \bibinfo {author}
  {\bibfnamefont {K.}~\bibnamefont {Pierloot}}, \bibinfo {author}
  {\bibfnamefont {F.}~\bibnamefont {Plasser}}, \bibinfo {author} {\bibfnamefont
  {M.}~\bibnamefont {Reiher}}, \bibinfo {author} {\bibfnamefont {A.~M.}\
  \bibnamefont {Sand}}, \bibinfo {author} {\bibfnamefont {I.}~\bibnamefont
  {Schapiro}}, \bibinfo {author} {\bibfnamefont {P.}~\bibnamefont {Sharma}},
  \bibinfo {author} {\bibfnamefont {C.~J.}\ \bibnamefont {Stein}}, \bibinfo
  {author} {\bibfnamefont {L.~K.}\ \bibnamefont {Sørensen}}, \bibinfo {author}
  {\bibfnamefont {D.~G.}\ \bibnamefont {Truhlar}}, \bibinfo {author}
  {\bibfnamefont {M.}~\bibnamefont {Ugandi}}, \bibinfo {author} {\bibfnamefont
  {L.}~\bibnamefont {Ungur}}, \bibinfo {author} {\bibfnamefont
  {A.}~\bibnamefont {Valentini}}, \bibinfo {author} {\bibfnamefont
  {S.}~\bibnamefont {Vancoillie}}, \bibinfo {author} {\bibfnamefont
  {V.}~\bibnamefont {Veryazov}}, \bibinfo {author} {\bibfnamefont
  {O.}~\bibnamefont {Weser}}, \bibinfo {author} {\bibfnamefont {T.~A.}\
  \bibnamefont {Wesołowski}}, \bibinfo {author} {\bibfnamefont {P.-O.}\
  \bibnamefont {Widmark}}, \bibinfo {author} {\bibfnamefont {S.}~\bibnamefont
  {Wouters}}, \bibinfo {author} {\bibfnamefont {A.}~\bibnamefont {Zech}},
  \bibinfo {author} {\bibfnamefont {J.~P.}\ \bibnamefont {Zobel}},\ and\
  \bibinfo {author} {\bibfnamefont {R.}~\bibnamefont {Lindh}},\ }\bibfield
  {title} {\bibinfo {title} {{OpenMolcas: From Source Code to Insight}},\
  }\href {https://doi.org/10.1021/acs.jctc.9b00532} {\bibfield  {journal}
  {\bibinfo  {journal} {J. Chem. Theor. Comput.}\ }\textbf {\bibinfo {volume}
  {15}},\ \bibinfo {pages} {5925} (\bibinfo {year} {2019})}\BibitemShut
  {NoStop}%
\bibitem [{\citenamefont {Aquilante}\ \emph {et~al.}(2020)\citenamefont
  {Aquilante}, \citenamefont {Autschbach}, \citenamefont {Baiardi},
  \citenamefont {Battaglia}, \citenamefont {Borin}, \citenamefont {Chibotaru},
  \citenamefont {Conti}, \citenamefont {De~Vico}, \citenamefont {Delcey},
  \citenamefont {Fdez.~Galv\'{a}n}, \citenamefont {Ferré}, \citenamefont
  {Freitag}, \citenamefont {Garavelli}, \citenamefont {Gong}, \citenamefont
  {Knecht}, \citenamefont {Larsson}, \citenamefont {Lindh}, \citenamefont
  {Lundberg}, \citenamefont {Malmqvist}, \citenamefont {Nenov}, \citenamefont
  {Norell}, \citenamefont {Odelius}, \citenamefont {Olivucci}, \citenamefont
  {Pedersen}, \citenamefont {Pedraza-González}, \citenamefont {Phung},
  \citenamefont {Pierloot}, \citenamefont {Reiher}, \citenamefont {Schapiro},
  \citenamefont {Segarra-Mart\'{i}}, \citenamefont {Segatta}, \citenamefont
  {Seijo}, \citenamefont {Sen}, \citenamefont {Sergentu}, \citenamefont
  {Stein}, \citenamefont {Ungur}, \citenamefont {Vacher}, \citenamefont
  {Valentini},\ and\ \citenamefont {Veryazov}}]{molcas2}%
  \BibitemOpen
  \bibfield  {author} {\bibinfo {author} {\bibfnamefont {F.}~\bibnamefont
  {Aquilante}}, \bibinfo {author} {\bibfnamefont {J.}~\bibnamefont
  {Autschbach}}, \bibinfo {author} {\bibfnamefont {A.}~\bibnamefont {Baiardi}},
  \bibinfo {author} {\bibfnamefont {S.}~\bibnamefont {Battaglia}}, \bibinfo
  {author} {\bibfnamefont {V.~A.}\ \bibnamefont {Borin}}, \bibinfo {author}
  {\bibfnamefont {L.~F.}\ \bibnamefont {Chibotaru}}, \bibinfo {author}
  {\bibfnamefont {I.}~\bibnamefont {Conti}}, \bibinfo {author} {\bibfnamefont
  {L.}~\bibnamefont {De~Vico}}, \bibinfo {author} {\bibfnamefont
  {M.}~\bibnamefont {Delcey}}, \bibinfo {author} {\bibfnamefont
  {I.}~\bibnamefont {Fdez.~Galv\'{a}n}}, \bibinfo {author} {\bibfnamefont
  {N.}~\bibnamefont {Ferré}}, \bibinfo {author} {\bibfnamefont
  {L.}~\bibnamefont {Freitag}}, \bibinfo {author} {\bibfnamefont
  {M.}~\bibnamefont {Garavelli}}, \bibinfo {author} {\bibfnamefont
  {X.}~\bibnamefont {Gong}}, \bibinfo {author} {\bibfnamefont {S.}~\bibnamefont
  {Knecht}}, \bibinfo {author} {\bibfnamefont {E.~D.}\ \bibnamefont {Larsson}},
  \bibinfo {author} {\bibfnamefont {R.}~\bibnamefont {Lindh}}, \bibinfo
  {author} {\bibfnamefont {M.}~\bibnamefont {Lundberg}}, \bibinfo {author}
  {\bibfnamefont {P.~{\AA}.}\ \bibnamefont {Malmqvist}}, \bibinfo {author}
  {\bibfnamefont {A.}~\bibnamefont {Nenov}}, \bibinfo {author} {\bibfnamefont
  {J.}~\bibnamefont {Norell}}, \bibinfo {author} {\bibfnamefont
  {M.}~\bibnamefont {Odelius}}, \bibinfo {author} {\bibfnamefont
  {M.}~\bibnamefont {Olivucci}}, \bibinfo {author} {\bibfnamefont {T.~B.}\
  \bibnamefont {Pedersen}}, \bibinfo {author} {\bibfnamefont {L.}~\bibnamefont
  {Pedraza-González}}, \bibinfo {author} {\bibfnamefont {Q.~M.}\ \bibnamefont
  {Phung}}, \bibinfo {author} {\bibfnamefont {K.}~\bibnamefont {Pierloot}},
  \bibinfo {author} {\bibfnamefont {M.}~\bibnamefont {Reiher}}, \bibinfo
  {author} {\bibfnamefont {I.}~\bibnamefont {Schapiro}}, \bibinfo {author}
  {\bibfnamefont {J.}~\bibnamefont {Segarra-Mart\'{i}}}, \bibinfo {author}
  {\bibfnamefont {F.}~\bibnamefont {Segatta}}, \bibinfo {author} {\bibfnamefont
  {L.}~\bibnamefont {Seijo}}, \bibinfo {author} {\bibfnamefont
  {S.}~\bibnamefont {Sen}}, \bibinfo {author} {\bibfnamefont {D.-C.}\
  \bibnamefont {Sergentu}}, \bibinfo {author} {\bibfnamefont {C.~J.}\
  \bibnamefont {Stein}}, \bibinfo {author} {\bibfnamefont {L.}~\bibnamefont
  {Ungur}}, \bibinfo {author} {\bibfnamefont {M.}~\bibnamefont {Vacher}},
  \bibinfo {author} {\bibfnamefont {A.}~\bibnamefont {Valentini}},\ and\
  \bibinfo {author} {\bibfnamefont {V.}~\bibnamefont {Veryazov}},\ }\bibfield
  {title} {\bibinfo {title} {{Modern quantum chemistry with [Open]Molcas}},\
  }\href {https://doi.org/10.1063/5.0004835} {\bibfield  {journal} {\bibinfo
  {journal} {J. Chem. Phys.}\ }\textbf {\bibinfo {volume} {152}},\ \bibinfo
  {pages} {214117} (\bibinfo {year} {2020})}\BibitemShut {NoStop}%
\bibitem [{\citenamefont {Iwahara}\ \emph {et~al.}(2017)\citenamefont
  {Iwahara}, \citenamefont {Vieru}, \citenamefont {Ungur},\ and\ \citenamefont
  {Chibotaru}}]{Iwahara2017}%
  \BibitemOpen
  \bibfield  {author} {\bibinfo {author} {\bibfnamefont {N.}~\bibnamefont
  {Iwahara}}, \bibinfo {author} {\bibfnamefont {V.}~\bibnamefont {Vieru}},
  \bibinfo {author} {\bibfnamefont {L.}~\bibnamefont {Ungur}},\ and\ \bibinfo
  {author} {\bibfnamefont {L.~F.}\ \bibnamefont {Chibotaru}},\ }\bibfield
  {title} {\bibinfo {title} {{Zeeman interaction and Jahn-Teller effect in the
  ${\mathrm{\ensuremath{\Gamma}}}_{8}$ multiplet}},\ }\href
  {https://doi.org/10.1103/PhysRevB.96.064416} {\bibfield  {journal} {\bibinfo
  {journal} {Phys. Rev. B}\ }\textbf {\bibinfo {volume} {96}},\ \bibinfo
  {pages} {064416} (\bibinfo {year} {2017})}\BibitemShut {NoStop}%
\bibitem [{\citenamefont {Iwahara}\ \emph {et~al.}(2018)\citenamefont
  {Iwahara}, \citenamefont {Vieru},\ and\ \citenamefont
  {Chibotaru}}]{Iwahara2018}%
  \BibitemOpen
  \bibfield  {author} {\bibinfo {author} {\bibfnamefont {N.}~\bibnamefont
  {Iwahara}}, \bibinfo {author} {\bibfnamefont {V.}~\bibnamefont {Vieru}},\
  and\ \bibinfo {author} {\bibfnamefont {L.~F.}\ \bibnamefont {Chibotaru}},\
  }\bibfield  {title} {\bibinfo {title} {Spin-orbital-lattice entangled states
  in cubic ${d}^{1}$ double perovskites},\ }\href
  {https://doi.org/10.1103/PhysRevB.98.075138} {\bibfield  {journal} {\bibinfo
  {journal} {Phys. Rev. B}\ }\textbf {\bibinfo {volume} {98}},\ \bibinfo
  {pages} {075138} (\bibinfo {year} {2018})}\BibitemShut {NoStop}%
\bibitem [{\citenamefont {Anderson}\ \emph {et~al.}(1999)\citenamefont
  {Anderson}, \citenamefont {Bai}, \citenamefont {Bischof}, \citenamefont
  {Blackford}, \citenamefont {Demmel}, \citenamefont {Dongarra}, \citenamefont
  {Du~Croz}, \citenamefont {Greenbaum}, \citenamefont {Hammarling},
  \citenamefont {McKenney},\ and\ \citenamefont {Sorensen}}]{laug}%
  \BibitemOpen
  \bibfield  {author} {\bibinfo {author} {\bibfnamefont {E.}~\bibnamefont
  {Anderson}}, \bibinfo {author} {\bibfnamefont {Z.}~\bibnamefont {Bai}},
  \bibinfo {author} {\bibfnamefont {C.}~\bibnamefont {Bischof}}, \bibinfo
  {author} {\bibfnamefont {S.}~\bibnamefont {Blackford}}, \bibinfo {author}
  {\bibfnamefont {J.}~\bibnamefont {Demmel}}, \bibinfo {author} {\bibfnamefont
  {J.}~\bibnamefont {Dongarra}}, \bibinfo {author} {\bibfnamefont
  {J.}~\bibnamefont {Du~Croz}}, \bibinfo {author} {\bibfnamefont
  {A.}~\bibnamefont {Greenbaum}}, \bibinfo {author} {\bibfnamefont
  {S.}~\bibnamefont {Hammarling}}, \bibinfo {author} {\bibfnamefont
  {A.}~\bibnamefont {McKenney}},\ and\ \bibinfo {author} {\bibfnamefont
  {D.}~\bibnamefont {Sorensen}},\ }\href@noop {} {\emph {\bibinfo {title}
  {{LAPACK} Users' Guide}}},\ \bibinfo {edition} {3rd}\ ed.\ (\bibinfo
  {publisher} {Society for Industrial and Applied Mathematics},\ \bibinfo
  {address} {Philadelphia, PA},\ \bibinfo {year} {1999})\BibitemShut {NoStop}%
\bibitem [{\citenamefont {Kotani}(1949)}]{Kotani1949}%
  \BibitemOpen
  \bibfield  {author} {\bibinfo {author} {\bibfnamefont {M.}~\bibnamefont
  {Kotani}},\ }\bibfield  {title} {\bibinfo {title} {{On the Magnetic Moment of
  Complex Ions. (I)}},\ }\href {https://doi.org/10.1143/JPSJ.4.293} {\bibfield
  {journal} {\bibinfo  {journal} {J. Phys. Soc. Jpn.}\ }\textbf {\bibinfo
  {volume} {4}},\ \bibinfo {pages} {293} (\bibinfo {year} {1949})}\BibitemShut
  {NoStop}%
\bibitem [{\citenamefont {Kotani}(1960)}]{Kotani1960}%
  \BibitemOpen
  \bibfield  {author} {\bibinfo {author} {\bibfnamefont {M.}~\bibnamefont
  {Kotani}},\ }\bibfield  {title} {\bibinfo {title} {{Properties of
  $d$-Electrons in Complex Salts. Part I Paramagnetism of Complex Salts}},\
  }\href {https://doi.org/10.1143/PTPS.14.1} {\bibfield  {journal} {\bibinfo
  {journal} {Prog. Theor. Phys. Suppl.}\ }\textbf {\bibinfo {volume} {14}},\
  \bibinfo {pages} {1} (\bibinfo {year} {1960})}\BibitemShut {NoStop}%
\bibitem [{\citenamefont {Abragam}\ and\ \citenamefont
  {Bleaney}(1970)}]{Abragam1970}%
  \BibitemOpen
  \bibfield  {author} {\bibinfo {author} {\bibfnamefont {A.}~\bibnamefont
  {Abragam}}\ and\ \bibinfo {author} {\bibfnamefont {B.}~\bibnamefont
  {Bleaney}},\ }\href@noop {} {\emph {\bibinfo {title} {{Electron Paramagnetic
  Resonance of Transition Ions}}}}\ (\bibinfo  {publisher} {Clarendon Press},\
  \bibinfo {address} {Oxford},\ \bibinfo {year} {1970})\BibitemShut {NoStop}%
\end{thebibliography}%

%

\end{document}